\documentclass[acmtog,screen,authorversion]{acmart} 

\acmSubmissionID{0209}

\usepackage{booktabs} 

\newcommand{\NEW}[1]{{\color{black}{#1}}} 

\usepackage{import} 
\graphicspath{{figs/}}

\usepackage[ruled,noend]{algorithm2e} 

\SetAlFnt{\small}
\SetAlCapFnt{\small}
\SetAlCapNameFnt{\small}
\SetAlCapHSkip{0pt}
\IncMargin{-\parindent}
\usepackage{ifpdf}
\usepackage{graphicx}
\ifpdf
  
\else
\fi
\usepackage{color}
\definecolor{cyan}{cmyk}{1,0,0,0}
\definecolor{darkgreen}{rgb}{0,0.5,0}
\definecolor{orange}{rgb}{1,0.5,0}
\definecolor{magenta}{cmyk}{0,1,0,0}
\definecolor{darkyellow}{cmyk}{0,0,0.75,0}
\definecolor{gray}{rgb}{0.8,0.8,0.8}

\newenvironment{tightitemize}{
\vspace{-1.5mm}
\begin{itemize}
  \setlength{\itemsep}{1pt}
  \setlength{\parskip}{2pt}
  \setlength{\parsep}{0pt}}{\end{itemize}

}
\usepackage{amsmath} 

\usepackage[toc,page]{appendix} 
\usepackage{wrapfig} 
\usepackage{comment}
\usepackage{bm}
\usepackage{cuted} 
\usepackage{lipsum} 
\usepackage{mathtools} 
\usepackage{algorithmicx}
\usepackage{algpseudocode}
\usepackage{nicefrac}

\usepackage{gensymb}
\usepackage{float}
\usepackage{soul}
\usepackage{array}
\usepackage{multirow}
\usepackage{hhline}
\usepackage{setspace}
\usepackage{nicefrac}

\algdef{SE}[DOWHILE]{Do}{doWhile}{\algorithmicdo}[1]{\algorithmicwhile\ #1}%

\makeatletter
\renewcommand{\ALG@beginalgorithmic}{\small}
\makeatother

\newcommand{\DELETE}[1]{} 
\newcommand{\IGNORE}[1]{}
\usepackage{datenumber}
\usepackage{calc}
\usepackage[mmddyyyy]{datetime}

\newcounter{datetoday}
\newcounter{diffyears}
\newcounter{diffmonths}
\newcounter{diffdays}

\newcommand{\difftoday}[3]{%
      \setmydatenumber{datetoday}{\the\year}{\the\month}{\the\day}%
      \setmydatenumber{diffdays}{#1}{#2}{#3}%
      \addtocounter{diffdays}{-\thedatetoday}%
      \ifnum\value{diffdays}>0
        \def\diffbefore{}%
        \def\diffafter{left}%
      \else
        \def\diffbefore{}%
        \def\diffafter{ago}%
        \setcounter{diffdays}{-\value{diffdays}}%
      \fi
      \setcounter{diffyears}{\value{diffdays}/365}%
      \setcounter{diffdays}{\value{diffdays}-365*\value{diffyears}}%
      \setcounter{diffmonths}{\value{diffdays}/30}%
      \setcounter{diffdays}{\value{diffdays}-30*\value{diffmonths}}%
      \diffbefore
      \ifnum\value{diffyears}=0
      \else
        \ifnum\value{diffyears}>1
            \thediffyears\space years,
        \else
            \thediffyears\space year,
        \fi
      \fi
      \ifnum\value{diffmonths}=0
      \else
        \ifnum\value{diffmonths}>1
            \thediffmonths\space months
        \else
            \thediffmonths\space month
        \fi
      \fi
      \ifnum\value{diffdays}=0
      \else
        \ifnum\value{diffdays}>1
            \thediffdays\space days
        \else
            \thediffdays\space day
        \fi
      \fi
      \diffafter
}

\def\thickhline{\noalign{\hrule height 1pt}}

\usepackage[ruled]{algorithm2e} 

\SetAlFnt{\small}
\SetAlCapFnt{\small}
\SetAlCapNameFnt{\small}
\SetAlCapHSkip{0pt}

\setcopyright{rightsretained}
\acmJournal{TOG}
\acmYear{2022}\acmVolume{41}\acmNumber{4}\acmArticle{133}\acmMonth{7} \acmDOI{10.1145/3528223.3530075}

\begin{document}

\title{Sparse Ellipsometry: Portable Acquisition of Polarimetric SVBRDF and Shape with Unstructured Flash Photography}

\author{Inseung Hwang}
\orcid{0000-0002-9971-1202}
\affiliation{%
  \institution{KAIST}
  \country{South Korea}}
\email{ishwang@vclab.kaist.ac.kr}

\author{Daniel S. Jeon}
\orcid{0000-0001-8433-6932}
\affiliation{%
  \institution{KAIST}
  \country{South Korea}}
\email{sjjeon@vclab.kaist.ac.kr}

\author{Adolfo Muñoz}
\orcid{0000-0002-8160-7159}
\affiliation{%
 \institution{Universidad de Zaragoza - I3A}
 \country{Spain}}
\email{adolfo@unizar.es}

\author{Diego Gutierrez}
\orcid{0000-0002-7503-7022}
\affiliation{%
  \institution{Universidad de Zaragoza - I3A}
  \country{Spain}}
\email{diegog@unizar.es}

\author{Xin Tong}
\orcid{0000-0001-8788-2453}
\affiliation{%
  \institution{Microsoft Research Asia}
  \country{China}}
\email{xtong@microsoft.com}

\author{Min H. Kim}
\orcid{0000-0002-5078-4005}
\affiliation{%
  \institution{KAIST}
  \country{South Korea}}
\email{minhkim@vclab.kaist.ac.kr}

\begin{abstract}
Ellipsometry techniques allow to measure polarization information of materials, requiring precise rotations of optical components with different configurations of lights and sensors. This results in cumbersome capture devices, carefully calibrated in lab conditions, and in very long acquisition times, usually in the order of a few days per object. Recent techniques allow to capture polarimetric spatially-varying reflectance information, but limited to a single view, or to cover all view directions, but limited to spherical objects made of a single  homogeneous material.
We present \textit{sparse ellipsometry}, a portable polarimetric acquisition method that captures both polarimetric SVBRDF and  3D shape simultaneously. Our handheld device consists of off-the-shelf, fixed optical components. Instead of days, the total acquisition time varies between twenty and thirty minutes per object. 
 We develop a complete polarimetric SVBRDF model that includes diffuse and specular components, as well as single scattering, and devise a novel polarimetric inverse rendering algorithm with data augmentation of specular reflection samples via generative modeling.
Our results show a strong agreement with a recent ground-truth dataset of captured polarimetric BRDFs of real-world objects.
\end{abstract}



  \begin{teaserfigure}
	\centering	
	\includegraphics[width=\linewidth]{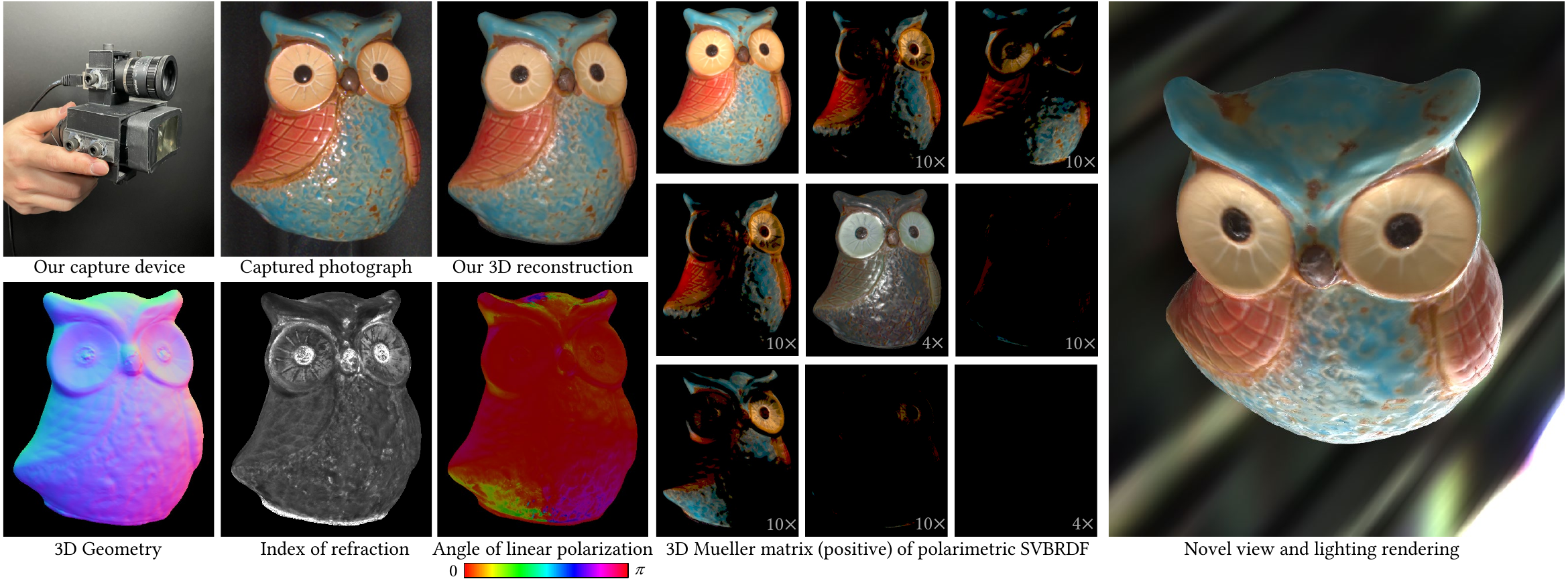}\\%
	\vspace{-3mm}%
	\caption[]{\label{fig:teaser}%
We propose the first sparse ellipsometry method that simultaneously captures both the polarimetric SVBRDF (including the \emph{3D Mueller matrix} and the index of refraction) and the 3D shape of real-world objects.
Different from traditional ellipsometry, our \emph{portable} acquisition device is made up of off-the-shelf, fixed optical components. Our sparse observations can be captured in minutes instead of days, allowing for accurate renderings of novel views under different illuminations.
}%
  \end{teaserfigure}

\maketitle



\section{Introduction}
\label{sec:intro}

Realistic modeling of the bidirectional reflectance distribution functions (BRDF) of real-world objects is a key aspect for physically based rendering.
The effect of scattering on the polarization state of light, however, has been traditionally ignored since it is mostly imperceptible to the human eye.
Nevertheless, polarization can be easily captured by an optical sensor, and provides useful information about the geometry and material properties of an object. 

Ellipsometry techniques use optical measurements to characterize how the interactions with a material change the polarization state of incident light~\cite{azzam2016stokes}. There are two main approaches, represented by two recent works. First, polarimetric spatially-varying reflectance information of real-world objects can be captured, but limited to a single view~\cite{Baek2018}; therefore, changes in viewpoint are not possible without visible distortions. Alternatively, polarimetric information can be measured from all view directions, but limited to spherical objects made of one single, homogeneous material~\cite{Baek2020}. Our work is the first to allow simultaneous capture of both polarimetric spatially-varying materials from all view directions, as well as arbitrary geometry.

Furthermore, while these recent approaches~\cite{Baek2018,Baek2020} require sophisticated tabletop setups and rotating optical equipment, combining polarization angles with different configurations of light and sensor, we design a portable device that combines an off-the-shelf polarization camera and flashlight with a linear polarizer, without the need for rotating elements. Using our device, we introduce a \textit{sparse ellipsometry} technique that allows us to capture both a complete linear polarimetric spatially-varying BRDF (SVBRDF) and the corresponding 3D geometry of a real-world object from a series of unstructured flash photographs; instead of days as with existing approaches, our method takes only between twenty and thirty minutes of acquisition time, while also yielding an accurate match with ground-truth data.

From the unstructured, sparse set of captured views, we recover polarimetric reflectance and  3D shape using an optimization algorithm that involves inverse rendering. Our polarimetric SVBRDF model includes not only diffuse and specular reflection, but subsurface single scattering as well, accurately matching ground-truth data recently captured with full ellipsometry techniques~\cite{Baek2020}. Since the sparse input data is often  insufficient to estimate \emph{per-vertex} specular parameters, specially for very narrow specular lobes, we devise a generative modeling strategy that augments the input with novel synthetic views. Our technique captures linear polarization, which corresponds to the first three components in the Stokes vector. Since we do not capture circular polarization, our scope is restricted to dielectric surfaces. 

Figure~\ref{fig:teaser} shows our portable device, along with a captured object, visualization of polarimetric data including the 3D Mueller matrix, and a novel view under different lighting conditions. Note that, even though all the views are captured only with the frontal flash lighting of our device, our inverse rendering technique is robust enough to render captured objects with arbitrary illumination even at grazing angles (see the rightmost image in Figure~\ref{fig:teaser}; please refer also to the supplemental video). 

In summary, our contributions are:
\begin{tightitemize}
	\item A method to capture both polarimetric SVBRDF and 3D geometry information simultaneously.
	\item A portable measurement device for polarimetric reflectance using off-the-shelf, fixed optics.
	\item A complete polarimetric SVBRDF model that includes diffuse and specular components, as well as single scattering. 
	\item A polarimetric inverse rendering algorithm that includes a generative modeling strategy for data augmentation.
\end{tightitemize}

\NEW{Our code is freely available for research purposes\footnote{https://github.com/KAIST-VCLAB/SparseEllipsometry.git}.}

\section{Related Work}

\label{sec:relatedwork}


\paragraph{Polarimetric BRDF models}
Polarimetric BRDF (pBRDF) models take into account changes in the polarization state of light as it interacts with the corresponding surface. Several pBRDF models~\cite{priest2000polarimetric,hyde2009geometrical,mojzik2016bipolar,jarabo2017bidirectional} define reflection as a mixture of a rough polarizing specular lobe (by accounting for Fresnel interactions in a microfacet model) and a depolarizing diffuse lobe. However, other works have shown diffuse (view independent) polarization effects~\cite{ellis1996polarimetric,sun2007statistical,maxwell1973bidirectional,Baek2020}, which can be explained by multiple subsurface scattering events (which depolarize light), followed by a Fresnel transmission from inside the object to the air, which partially polarizes light~\cite{Atkinson2006}.

Recent pBRDF models consider this diffuse polarizing effect~\cite{kadambi2015polar,cui2017polarimetric,Baek2018}, and even leverage it to estimate surface normals.
Our pBRDF model includes both polarizing diffuse and specular lobes, plus a new single scattering lobe, thus providing an accurate match against \textit{measured} ground-truth data~\cite{Baek2020}. 

%
%

\paragraph{Polarimetric imaging} Polarization information has been used for a wide range of imaging purposes, such as dehazing~\cite{Liu2015}, elimination of specular highlights~\cite{yang2016method}, or new image editing techniques~\cite{delMolino2019polarization}. However, separating specular and diffuse polarization from reflected light is an open research problem due to the complex changes in polarization caused by surface scattering. 

Some works take advantage of the polarizing effect of the specular component to estimate reflectance parameters and/or geometry. \citet{Ma2007} separate diffuse reflectance from specular components using polarized gradient illumination. More complete appearance models, including per-pixel diffuse and specular albedos, global roughness, and normal map have been proposed: \citet{Ghosh2008} extract layers of skin reflectance using polarization-difference images and data-driven techniques, for the particular case of faces. Other works leverage circularly polarized light~\cite{Ghosh2010}, polarized gradient illumination~\cite{Ghosh2011}, outdoors sky light~\cite{Riviere2017}, or including cross-polarization filters in cameras~\cite{Riviere2020}. However, the high-frequency behavior and angular dependency of the specular reflection require multiple views and/or large-scale lighting setups to be able to find and leverage the specular component.

Another strategy considers the polarization effect of the diffuse reflection state as a function of the azimuth angle of the surface normal, given the Fresnel transmission from the object to the air. This has been used for geometry reconstruction (SfP, structure from polarization), estimating geometry and normals from single view polarimetric images~\cite{Miyazaki2003,Atkinson2006,Huynh2013,Tozza2017} or multiple views~\cite{cui2017polarimetric,Zhu2019,Cui2019,Zhao2020}. Other methods have used deep neural networks to reconstruct both normals and spatially-varying material properties (i.e., SVBRDF) based on the same diffuse polarization cues~\cite{Ba2020,Lei2021,Deschaintre2021}. Since light depolarizes as it enters the object and given that the diffuse reflection is low frequency, these methods work even under unknown lighting conditions, while generally requiring simpler setups. However, the specular component of reflectance is not properly accounted for, which may lead to errors.
 Moreover, these existing methods leverage polarization information mainly to estimate diffuse albedo and normal maps, but they do not acquire complete polarimetric reflectance information, such as full Mueller matrices and indices of refraction.

\citet{Baek2018} introduced a pBRDF model as a linear combination of polarizing diffuse and specular components. Since then, there have been works on structure from polarization methods that account for both polarizing effects~\cite{Fukao2021,Ding2021}. However, these methods require a full Mueller matrix per pixel, which involves multiple shots changing the polarization state of the incident light. This requires rotation by hardware, which translates into heavy, sophisticated capture systems. These complex Mueller matrix measurements have also been used for separating components of light transport~\cite{Baek2021} and for measuring pBRDFs~\cite{Baek2020}. Our work also considers the effect of both polarizing diffuse and specular components but, in contrast, it only requires a single Stokes vector per pixel. Consequently, it does not require a heavy polarimetric control of external light sources, leading to a compact hand-held capture device.
In contrast, our portable design allows us to integrate per-pixel Stokes vectors observations from multiple views, enabling for the first time simultaneous acquisition of full polarimetric SVBRDF (including 3D Mueller matrix and index of refraction) and 3D shape.

\paragraph{3D geometry and SVBRDF acquisition}
Traditionally, simultaneous acquisition of geometry and SVBRDF required large-scale specialized setups, such as light stages including polarized light sources, multiple cameras and/or projectors for structured lighting~\cite{Ma2007,Ghosh2008,Ghosh2010}. These setups have become progressively simpler, requiring 
\NEW{a professional setup with multiple cameras and polarized lights}~\cite{Ghosh2011}, a rotating LED pattern for spherical harmonics~\cite{Tunwattanapong2013}, controlled front LED lighting with multiple cameras~\cite{Gotardo2018}, or spherical LED lighting with a single camera~\cite{kampouris2018diffuse}.

Other works focus on further simplifying the hardware setup for the capture, relying on photometric stereo~\cite{Zhou2013}, calibrated environments~\cite{Oxholm2014}, RGB-D information~\cite{Wu2016}, video sequences~\cite{Xia2016}, 
\NEW{unstructured flash photographies~\cite{Nam2018}, or pretrained structured LED illumination~\cite{ma2021free}.}
More recently, NeRF networks~\cite{mildenhall2020nerf} are able to reconstruct geometry and appearance parameters from images in arbitrary lighting condition~\cite{Srinivasan2021,boss2021nerd}. Our capture system also consists on a simple and hand-held setup that includes a flashlight, a polarizer, and polarization camera. Different from existing works, however, we not only acquire 3D geometry and SVBRDF, but also polarization.
\section{Polarimetric Image Formation}
\label{sec:modeling}

\subsection{Background on Polarization}
\label{sec:method-background}

The Stokes vector $\mathbf{s}={{\left[ {{s}_{0}},{{s}_{1}},{{s}_{2}},{{s}_{3}} \right]}^{\top }}$ is a four-dimensional quantity that describes the polarization state of light. 
$s_0$ represents total radiance, while $s_1$, $s_2$, and $s_3$ are the polarized components defined by the degree of polarization $\psi$, polarization angle $\xi$, ellipticity angle~$\zeta$ as $\mathbf{s}=\left[ {{s}_{0}},{{s}_{0}}\psi \cos 2\zeta \cos 2\xi ,{{s}_{0}}\psi \cos 2\zeta \sin 2\xi ,{{s}_{0}}\psi \sin 2\zeta  \right]$. The  $s_1$,  $s_2$ and  $s_3$ components describe horizontal, linearly diagonal, and circular polarization, respectively.

The Mueller matrix $\mathbf{M}$ represents a transformation of the polarization state of light represented by a Stokes vector: $\mathbf{s}_a=\mathbf{M}\mathbf{s}_b$, where $\mathbf{s}_{a,b}$ are the Stokes vectors before and after any transformation event, such as a change of coordinates, reflection, transmission or filtering.

A change on the polarization state after a reflection or transmission on an interface depends on Fresnel equations, which indicates that light polarized along the axis parallel to the plane of incidence is affected differently than light polarized perpendicular to that plane of incidence. Therefore, before the interaction, we need to align one of the axis (the $y$-axis in our case) to be parallel to the plane of incidence, which requires a rotation of the local frame \NEW{(see Section~\ref{sec:appearance-model})}. Specifically, the counterclockwise rotation with angle~$\vartheta$ is represented by the following matrix $\mathbf{C}$ as 
\begin{equation}
	\label{eq:coord_matrix}
\mathbf{C}\left( \vartheta  \right)=\left[ \begin{matrix}
	1 & 0 & 0 & 0  \\
	0 & \cos 2\vartheta  & \sin 2\vartheta  & 0  \\
	0 & -\sin 2\vartheta  & \cos 2\vartheta  & 0  \\
	0 & 0 & 0 & 1  \\
\end{matrix} \right].
\end{equation}

Once in the adequate local frame, transmission and reflection of light can be represented by the Fresnel matrix $\mathbf{F}$:
\begin{equation}
	\label{eq:fresnel_matrix}
	{{\mathbf{F}}^{F \in \{T, R \}}}=\left[ \begin{matrix}
		\frac{{{F}^{\bot }}+{{F}^{\parallel }}}{2} & \frac{{{F}^{\bot }}-{{F}^{\parallel }}}{2} & 0 & 0  \\
		\frac{{{F}^{\bot }}-{{F}^{\parallel }}}{2} & \frac{{{F}^{\bot }}+{{F}^{\parallel }}}{2} & 0 & 0  \\
		0 & 0 & \sqrt{{{F}^{\bot }}{{F}^{\parallel }}}\cos \delta  & \sqrt{{{F}^{\bot }}{{F}^{\parallel }}}\sin \delta   \\
		0 & 0 & -\sqrt{{{F}^{\bot }}{{F}^{\parallel }}}\sin \delta  & \sqrt{{{F}^{\bot }}{{F}^{\parallel }}}\cos \delta   \\
	\end{matrix} \right],
\end{equation}
where $F \in \{T, R \}$ represents the Fresnel transmission ($T$) or reflection ($R$) coefficients, and $\delta$ is the retardation (delay) phase shift  ($0$~when the incident angle is larger than the Brewster angle, $\pi$ otherwise).
${{F}^{\bot }}$ and ${{F}^{\parallel }}$ are the Fresnel coefficients for the perpendicular and parallel components with respect to the plane of incidence (see the work of~\citet{wilkie2012polarised} for a complete description of polarized light). 
In the following $F$ will refer to $T$ or $R$, depending on the specific interaction (transmission or reflection) being considered. For convenience, notations are summarized in Table~\ref{tab:notations}.

\subsection{Polarimetric Reflectance Model}
\label{sec:appearance-model}

The light transport of polarized light can be defined in terms of the Stokes vector and the pBRDF as
\begin{equation}
	\label{eq:light_transport}
	{{\mathbf{s}}_{o}}=S \mathbf{P}\left(\bm{\omega}_{i},\bm{\omega}_{o}\right){{\mathbf{s}}_{i}}.
\end{equation}
where $S=\frac{\left( \mathbf{n}\cdot {{\mathbf{\omega }}_{i}} \right)}{d^2}$ is the shading term including light attenuation, $d$~is the distance between the light and the surface and $\mathbf{P}\left(\bm{\omega}_{i},\bm{\omega}_{o}\right)$ is the pBRDF that yields a Mueller matrix for a specific incoming and outgoing direction.

Recent pBRDF models follow microfacet theory~\cite{torrance1967theory,Cook:1982:BRDF,VGROOVE:SIGA:2018} and separate reflectance into diffuse and specular terms~\cite{Baek2018,Baek2021}. However, our experiments have shown that the combination of those two lobes does not match captured datasets for polarimetric appearance~\cite{Baek2020}.
We thus introduce a new additional term that takes into account \textit{single scattering} effects; our resulting model provides a better fit against captured polarization data  (see Section~\ref{sec:result-scattering}). 

Our new polarimetric pBRDF model can then be expressed as
\begin{equation}
	\label{eq:diffuse_and_specular}
	\mathbf{P}={{\mathbf{P}}^{d}}+{{\mathbf{P}}^{s}}+{{\mathbf{P}}^{ss}},
\end{equation}
where ${\mathbf{P}}^{d}$, ${\mathbf{P}}^{s}$ and ${\mathbf{P}}^{ss}$ are the pBRDF lobes of diffuse reflection, specular reflection, and single scattering, respectively. The dependency on $\bm{\omega}_{i}$ and $\bm{\omega}_{o}$ has been omitted for the sake of brevity.

\begin{figure}[]
	\centering	\includegraphics[width=\linewidth]{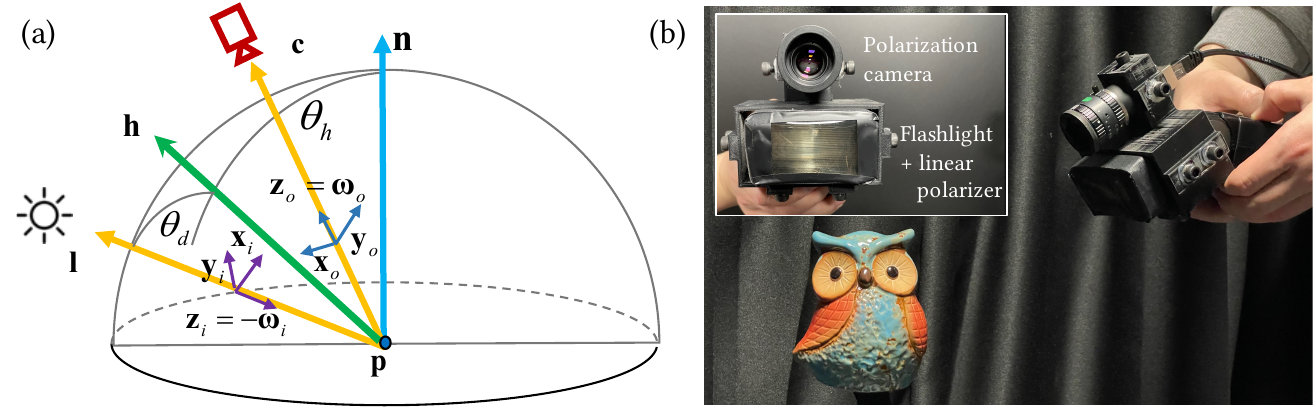}%
	\vspace{-3mm}%
	\caption[]{\label{fig:coordinates}%
 (a) Our polarimetric coordinate system. (b) Our handheld polarimetric imaging setup consisting of a polarimetric camera and a photographic flashlight.
	}%
	\vspace{-3mm}
\end{figure}

\paragraph{Local polarization frame} 
Our coordinate system is defined by a local frame with three orthonormal vectors, where the $z$-axis follows the propagation of light, 
\NEW{the $y$-axis is aligned with the camera up vector for outgoing light, and with the orthogonal direction of the horizontal linear polarization filter on the light for incoming light,}
and the $x$-axis is perpendicular to both (see Figure~\ref{fig:coordinates}a). 
This is different from previous work~\cite{Baek2018} in that we do not need to flip axes for incoming and outgoing light; instead, the frame is rotated so that its $y$-axis is aligned to the plane of incidence, thus matching the expected orientation for Fresnel interactions. 
The plane of incidence is defined by the surface normal $\mathbf{n}$ for the diffuse lobe and by the halfway vector $\mathbf{h}$ for the specular and single scattering lobes. We define the polarimetric azimuth angles for incident and exitant light corresponding to these two planes, respectively: $\phi_{\{i,o\}} = \tan^{-1}((\mathbf{n}\cdot {{\mathbf{y}}_{\{i,o\}}})/ (\mathbf{n}\cdot {\mathbf{x}_{\{i,o\}}}) )$ and ${{\varphi }_{\{i,o\}}} = \tan^{-1} ((\mathbf{h}\cdot {\mathbf{y}_{\{i,o\}}})/(\mathbf{h}\cdot {{\mathbf{x}}_{\{i,o\}}}) )$. Therefore, the corresponding rotation angles required to match the incident and exitant frames to the interaction plane are $\phi^{\llcorner}_{\{i,o\}} = \phi_{\{i,o\}} - \pi/2$ and $\varphi^{\llcorner}_{\{i,o\}} = \varphi_{\{i,o\}} - \pi/2$, respectively. 

\paragraph{Diffuse and specular terms}
We adopt the diffuse and specular terms from state-of-the-art pBRDF models~\cite{Baek2018,Baek2021}, which we briefly summarize here for completeness. \textit{Polarized diffuse reflection} is formulated as 
\begin{equation}
	\label{eq:diffuse_transport}
	{{\mathbf{P}}^{d}}={{\mathbf{C}}_{n\to o}}\left(-{\phi^{\llcorner}_{o}} \right){{\mathbf{F}}^{T}}\left( {{\theta }_{o}};\eta  \right)\mathbf{D}\left( \rho_d  \right){{\mathbf{F}}^{T}}\left( {{\theta }_{i}};\eta  \right){{\mathbf{C}}_{i\to n}}\left( {\phi^{\llcorner}_{i}}\right),
\end{equation}
where $\mathbf{D}$ is the depolarization matrix with diffuse albedo $\rho_d$. In Mueller matrix form, ${{\mathbf{P}}^{d}}$ is
\begin{equation}
	\label{eq:diffuse_mueller}
	{{\mathbf{P}}^{d}}=\rho_d \left[ \begin{matrix}
		T_{o}^{+}T_{i}^{+} & -T_{o}^{+}T_{i}^{-}{{\beta }_{i}} & -T_{o}^{+}T_{i}^{-}{{\alpha }_{i}} & 0  \\
		-T_{o}^{-}T_{i}^{+}{{\beta }_{o}} & T_{o}^{-}T_{i}^{-}{{\beta }_{i}}{{\beta }_{o}} & T_{o}^{-}T_{i}^{-}{{\alpha }_{i}}{{\beta }_{o}} & 0  \\
		-T_{o}^{-}T_{i}^{+}{{\alpha }_{o}} & T_{o}^{-}T_{i}^{-}{{\alpha }_{o}}{{\beta }_{i}} & T_{o}^{-}T_{i}^{-}{{\alpha }_{o}}{{\alpha }_{i}} & 0  \\
		0 & 0 & 0 & 0  \\
	\end{matrix} \right],
\end{equation}
where $\alpha_{\{i,o\}}$ and $\beta_{\{i,o\}}$ indicate $\sin (2\phi_{\{i,o\}})$ and $\cos (2\phi_{\{i,o\}} )$, respectively. 
Here $T^ + _{\{i,o\}}$ and $T^ - _{\{i,o\}}$ are computed from incident/exitant Fresnel transmission coefficients,  $(T^ \bot _{\{i,o\}} + T^\parallel _{\{i,o\}} )/{2}$ and $(T^ \bot _{\{i,o\}} - T^\parallel _{\{i,o\}} )/2$. 

\textit{Polarized specular reflection}, on the other hand, is described as a single-bounce reflection on the microfacets as 
%
\begin{align}
	\label{eq:specular_transport}
{{\mathbf{P}}^{s}}= \kappa_{s}{{\mathbf{C}}_{h\to o}}\left(-{\varphi^{\llcorner}_{o}} \right){{\mathbf{F}}^{R}}\left( {{\theta }_{d}};\eta  \right){{\mathbf{C}}_{i\to h}}\left({\varphi^{\llcorner}_{i}} \right),
\end{align}
where $\kappa_{s}={{\rho}_{s}}\frac{D\left( {{\theta }_{h}};{{\sigma }_{s}} \right)G\left( {{\theta }_{i}},{{\theta }_{o}};{{\sigma }_{s}} \right)}{4\left( \mathbf{n}\cdot \bm{\omega}_{i} \right)\left( \mathbf{n}\cdot \bm{\omega}_{o} \right)}$ is the specular reflection term,
$D$~is the normal GGX distribution function~\cite{walter2007microfacet}, ${\sigma }_{s}$~is surface roughness, $G$ is Smith's geometric attenuation function of shadowing/masking~\cite{Heitz2014Microfacet}, and ${\rho}_{s}$ is the specular albedo.
Assuming that our target surface is dielectric, specular reflection can be considered monochromatic.
The specular lobe ${{\mathbf{P}}^{s}}$ can be expressed in Mueller matrix form as
\begin{align}
	\label{eq:specular_mueller}
&	{{\mathbf{P}}^{s}}=	\\ \nonumber
& \resizebox{1.0\columnwidth}{!}{
	\mbox{\fontsize{10}{12}\selectfont $	
	\kappa_{s} \left[ \begin{matrix}
		{{R}^{+}} & -{{R}^{-}}{{\gamma }_{i}} & -{{R}^{-}}{{\chi }_{i}} & 0  \\
		-{{R}^{-}}{{\gamma }_{o}} & {{R}^{+}}{{\gamma }_{i}}{{\gamma }_{o}}+{{R}^{\times }}{{\chi }_{i}}{{\chi }_{o}}\cos \delta  & {{R}^{+}}{{\chi }_{i}}{{\gamma }_{o}}-{{R}^{\times }}{{\gamma }_{i}}{{\chi }_{o}}\cos \delta  & {{R}^{\times }}{{\chi }_{o}}\sin \delta   \\
		-{{R}^{-}}{{\chi }_{o}} & {{R}^{+}}{{\gamma }_{i}}{{\chi }_{o}}-{{R}^{\times }}{{\chi }_{i}}{{\gamma }_{o}}\cos \delta  & {{R}^{+}}{{\chi }_{i}}{{\chi }_{o}}+{{R}^{\times }}{{\gamma }_{i}}{{\gamma }_{o}}\cos \delta  & -{{R}^{\times }}{{\gamma }_{o}}\sin \delta   \\
		0 & -{{R}^{\times }}{{\chi }_{i}}\sin \delta  & {{R}^{\times }}{{\gamma }_{i}}\sin \delta  & {{R}^{\times }}\cos \delta 
	\end{matrix} \right],	
	$}}
\end{align}
where $\chi_{\{i,o\}}$ and $\gamma_{\{i,o\}}$ denote $\sin(2\varphi_{\{i,o\}})$ and $\cos(2\varphi_{\{i,o\}})$, respectively. For a dielectric surface
$\cos\delta=-1$ when the incident angle is less than the Brewster angle, and $1$ otherwise.


\paragraph{Single scattering term}
\label{sec:single-scattering-model}
We extend existing pBRDF models by incorporating a new \textit{polarimetric single scattering} term, which yields a better match with captured data.
\begin{figure}[tpb]
	\centering
	\vspace{-3mm}	
	\includegraphics[width=\linewidth]{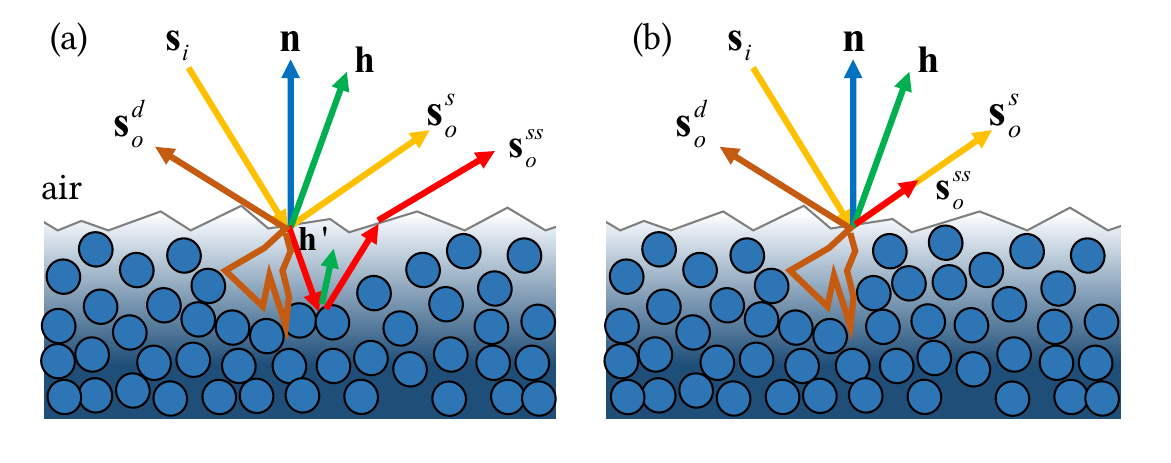}%
	\vspace{-6mm}
\caption{\label{fig:single-scattering}%
(a) Geometry of our polarimetric scattering model. 
\NEW{The $\mathbf{s}^{s}_{o}$ yellow vector represents the specular term, the brown vector ~$\mathbf{s}^{d}_{o}$ represents the diffuse term, and the red vector~$\mathbf{s}^{ss}_{o}$ represents our additional single scattering term. }
(b) Our simplified scattering model for polarimetric inverse rendering.}
	\vspace{-3mm}		
\end{figure}
In our model, the path of light for a single scattering event includes a Fresnel transmission into the object, a scattering reflection, and a second Fresnel transmission back into the air ($\mathbf{s}^{ss}_{o}$, see red vectors in Figure~\ref{fig:single-scattering}a). Different from specular reflection, color can change according to the albedo of the medium $\rho_{ss}$, while its angular distribution is a combination of the roughness of the surface (which affects Fresnel transmission) and the phase function $p\left(\bm{\omega}_i, \bm{\omega}_o\right)$. Our single-scattering term can thus be modeled as
\begin{align}
	\label{eq:single_scattering_transport}
{{\mathbf{P}}^{ss}} = &{{\mathbf{C}}_{n \to o}}\left(-{\phi^{\llcorner}_{o}} \right){{\mathbf{F}}^T}\left( {{\theta _o} ;\eta} \right) {{\mathbf{C}}_{h' \to n}}\left(-{\varphi^{\llcorner'}_{o}} \right) {{\mathbf{F}}^R}\left( {{\theta'_d}};\eta_p \right)\\ \nonumber 
& \cdot {r _{ss}} {{\mathbf{C}}_{n \to h'}}\left( {\varphi^{\llcorner'}_{i}} \right){{\mathbf{F}}^T}\left( {{\theta _i};\eta } \right){{\mathbf{C}}_{i \to n}}\left( \phi^{\llcorner}_i \right),
\end{align}
where ${\varphi^{\llcorner'}_{\{i,o\}}}$ is the rotation angle w.r.t. the medium's microfacet $\mathbf{h}'$ vector for incoming/outgoing rays (Figure~\ref{fig:single-scattering}a), and $\eta$ and $\eta_p$ are the indices of refraction for the surface and the single scattering interactions, respectively.
The parameter $r _{ss}$ represents the single-scattering BRDF \cite{hanrahan1993reflection} as:
\begin{equation}
	\label{eq:single_scattering_brdf}
	r _{ss} = \rho_{ss} p\left(\bm{\omega}'_i, \bm{\omega}'_o \right)\frac{1}{\mathbf{h}' \cdot \bm{\omega}'_i + \mathbf{h}' \cdot\bm{\omega}'_o },
\end{equation}
where directions $\bm{\omega}'_i$ and $\bm{\omega}'_o$ are deterministically calculated using Snell's Law, and the phase function $p\left(\bm{\omega}'_i, \bm{\omega}'_o \right)$ is the angular distribution that may be represented by Henyey-Greenstein's model~\cite{henyey1941diffuse}. 

While physically accurate, the model defined in Equation~\eqref{eq:single_scattering_transport} is ill-posed and not suitable for inverse rendering, since the single scattering event is occluded from direct view after a refractive event with an unknown index of refraction. We therefore introduce a simplified model, approximating the physically-based behavior of our single scattering component and making it practical for imaging and inverse rendering applications. We make the following observations:

\begin{itemize}  
  \item Single scattering presents a polarization state similar to the polarization state of specular reflections~\cite{Ghosh2008}.
   \item The angular distribution (roughness) of single scattering comes from the combination of the roughness of both Fresnel transmissions and the phase function. 
   \item While specular reflection does not change color, single scattering modifies color according to the albedo and extinction of the medium.
\end{itemize}

From these observations, our single scattering component can be represented as an extension of the specular reflection term, with independent roughness and albedo parameters:
\begin{align}
	\label{eq:single_scattering_transport2}
&{{\mathbf{P}}^{ss}}= {{\kappa}_{ss}} {{\mathbf{C}}_{h\to o}}\left(-{\varphi^{\llcorner}_{o}} \right){{\mathbf{F}}^{R}}\left( {{\theta }_{d}};\eta  \right){{\mathbf{C}}_{i\to h}}\left({\varphi^{\llcorner}_{i}} \right),
\end{align}
where $\kappa_{ss}={{\rho}_{ss}}\frac{D\left( {{\theta }_{h}};{{\sigma }_{ss}} \right)G\left( {{\theta }_{i}},{{\theta }_{o}};{{\sigma }_{ss}} \right)}{4\left( \mathbf{n}\cdot \bm{\omega}_{i} \right)\left( \mathbf{n}\cdot \bm{\omega}_{o} \right)}$ describes the single scattering lobe, and ${\sigma }_{ss}$ represents surface roughness for single scattering.
Note that the single scattering albedo ${\rho}_{ss}$ is a colored vector, as opposed to the single-value specular component.


Mueller matrix representation for the single scattering lobe is
\begin{align}
	\label{eq:single_scattering_mueller}
&{{\mathbf{P}}^{ss}}=  \\ \nonumber
& \resizebox{1.0\columnwidth}{!}{
	\mbox{\fontsize{10}{12}\selectfont $		
	\kappa_{ss} \left[ \begin{matrix}
		{{R}^{+}} & -{{R}^{-}}{{\gamma }_{i}} & -{{R}^{-}}{{\chi }_{i}} & 0  \\
		-{{R}^{-}}{{\gamma }_{o}} & {{R}^{+}}{{\gamma }_{i}}{{\gamma }_{o}}+{{R}^{\times }}{{\chi }_{i}}{{\chi }_{o}}\cos \delta  & {{R}^{+}}{{\chi }_{i}}{{\gamma }_{o}}-{{R}^{\times }}{{\gamma }_{i}}{{\chi }_{o}}\cos \delta  & {{R}^{\times }}{{\chi }_{o}}\sin \delta   \\
		-{{R}^{-}}{{\chi }_{o}} & {{R}^{+}}{{\gamma }_{i}}{{\chi }_{o}}-{{R}^{\times }}{{\chi }_{i}}{{\gamma }_{o}}\cos \delta  & {{R}^{+}}{{\chi }_{i}}{{\chi }_{o}}+{{R}^{\times }}{{\gamma }_{i}}{{\gamma }_{o}}\cos \delta  & -{{R}^{\times }}{{\gamma }_{o}}\sin \delta   \\
		0 & -{{R}^{\times }}{{\chi }_{i}}\sin \delta  & {{R}^{\times }}{{\gamma }_{i}}\sin \delta  & {{R}^{\times }}\cos \delta  
	\end{matrix} \right] 
$}},
\end{align}
where $\chi_{\{i,o\}}$, $\gamma_{\{i,o\}}$ and $\cos\delta$ are the same as in specular reflection. 
\subsection{Portable Acquisition Setup}
\label{sec:image-acquisition}\label{sec:handheld-setup}
Capturing full polarimetric appearance information of an object requires exhaustive sampling in both traditional ellipsometry~\cite{azzam1978photopolarimetric} and image-based ellipsometry~\cite{Baek2020} methods, which usually takes between two and five days. In addition, the process needs a complex tabletop setup, with multiple rotating polarization filters and retarders. We aim for a more usable, efficient and practical approach, for which we leverage our (spatially varying) pBRDF model (Section~\ref{sec:appearance-model}); while simpler and easier to approximate, it provides a very good match with full polarimetric measured data~\cite{Baek2020}, as our results will show.

We propose a simple setup that consists of a polarimetric camera with \textit{fixed} linear polarization filters at four different orientations (0$^\circ$, 45$^\circ$, 90$^\circ$, and 135$^\circ$), and a photographic flashlight with a fixed linear polarization filter at 0$^\circ$ angle.  As shown in Figure~\ref{fig:coordinates}(b), the light and the camera are mounted close together in a near-coaxial setup (around $3.5^\circ$, but it needs to be calibrated), leading to a small enough device for handheld acquisition.  Our input is a set of multiple RAW images including four different polarization angles. The intensity of the flash varies randomly for each captured image (between \{1/4, 1/8, 1/16\} of its maximum intensity), so per-vertex HDR radiance can be recovered~\cite{Debevec:1997:RHD}.

Instead of rotating optical components in a complex, fixed tabletop setup, we leverage the portable form factor of our capture device and develop a \textit{single-view} image formation model (Section~\ref{sec:image-formation model}), and apply this model in the optimization of multiple unstructured observations (Section~\ref{sec:multiview}). %

\subsection{Image Formation Model}
\label{sec:image-formation model}

Section~\ref{sec:appearance-model} introduces our complete pBRDF model. We now customize such a model to formulate light transport specifically for our acquisition system where the camera and flashlight are narrowly placed and oriented (see Section~\ref{sec:handheld-setup}), and represent the physical magnitudes of our pBRDF model as observations over the four captured images.

This narrow setup has been used in a recent study~\cite{Baek2018}. It is clearly beneficial in polarimetric imaging because it enables many simplifications of polarized interactions with surfaces. Our hardware design takes the advantage of this simplification by inheriting this near-coaxial setup. However, the approach by \citet{Baek2018} is based on ellipsometry, exhaustively capturing sixteen combinations of incident/exitant polarization states as input, while our approach only requires four combinations (the four linear filters of the camera and only one for the light source).

\begin{figure}[tpb]
	\centering	\includegraphics[width=0.95\linewidth]{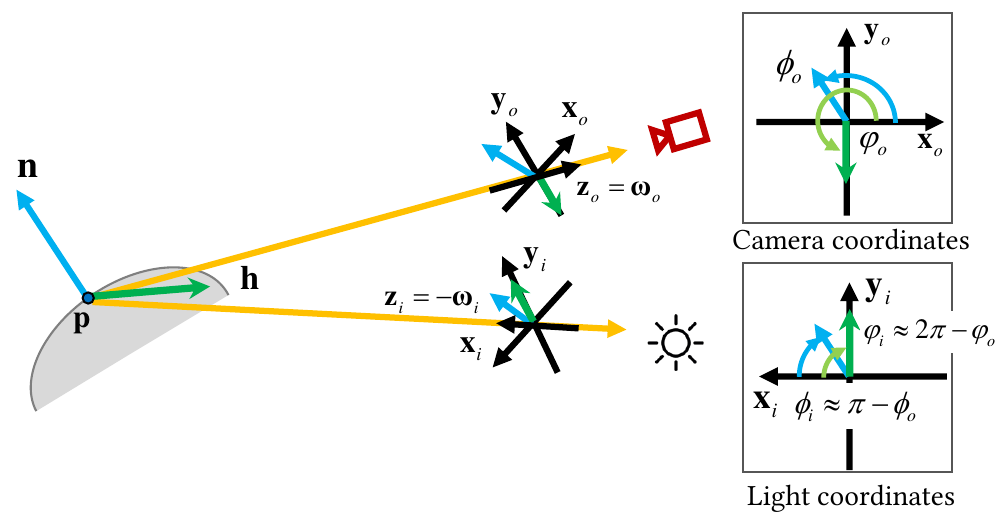}%
	\vspace{-3mm}%
	\caption[]{\label{fig:azimuth_angle}%
		The geometry of our portable setup is narrow, with the flashlight slightly below the camera sensor. This establishes several geometrical relationships between the light polarization frame, the camera polarization frame, surface normal $n$, and halfway vector $h$. This affects the relationship between the polarimetric azimuths for the diffuse lobe ($\phi_{i}\approx\pi-\phi_{o}$) with respect to the normal. Also, the projection of the halfway vector into the incident polarization plane points towards the exitant frame and vice versa, therefore establishing a relation between the corresponding polarization azimuths ($\varphi_{i}\approx 2\pi-\varphi_{o}$). 
	}%
	\vspace{-4mm}
\end{figure}

Some of these simplifications are purely geometrical: as light and camera rays are only separated by approximately $3.5^\circ$, we can assume $\bm{\omega}_i \approx \bm{\omega}_o$ and, as the optical elements are equally oriented, incident and exitant local polarization frames are also related as $\phi_{i}\approx \pi-\phi_{o}$ and $\varphi_{i}\approx 2\pi-\varphi_{o}$ so therefore $\alpha_{i}\approx-\alpha_{o}$, $\beta_{i}\approx\beta_{o}$,$\chi_{i}\approx-\chi_{o}$ and $\gamma_{i}\approx\gamma_{o}$. Furthermore, incident angle is by definition below Brewster angle so $\cos\delta=-1$ for both specular and single scattering terms. The geometrical relations defined by this narrow geometry are illustrated in Figure~\ref{fig:azimuth_angle}.

Last, the parallel and perpendicular components of the Fresnel reflection become close to each other ($R^{\parallel}\approx R^{\bot}$), so that $R^{-}\approx 0$ and $R^{+}\approx R^{\times}$ and light from the diffuse interaction gets depolarized through multiple subsurface scattering so the degree of polarization of incoming light $T^{-}_{o }T^{-}_{i}\approx 0$.



Adding the three lobes defined in Equations~\eqref{eq:diffuse_mueller}, \eqref{eq:specular_mueller}, and \eqref{eq:single_scattering_mueller}, and taking into account these simplifications, our full pBRDF as a Mueller matrix can be obtained as 
\begin{equation}
	\label{eq:coaxial_mueller}
	\mathbf{P}\approx 
\resizebox{0.92\linewidth}{!}{
	\mbox{\fontsize{10}{12}\selectfont $
	\left[ \begin{matrix}
		\rho_{d} T_{{}}^{+}T_{{}}^{+}+{{\kappa }_{s}}{{R}^{+}}+{{\kappa }_{ss}}{{R}^{+}} & -\rho_{d} T_{{}}^{-}T_{{}}^{+}{{\beta }_{{}}} & \rho_{d} T_{{}}^{-}T_{{}}^{+}{{\alpha }_{{}}} & 0  \\
		-\rho_{d} T_{{}}^{-}T_{{}}^{+}{{\beta }_{{}}} & {{\kappa }_{s}}{{R}^{+}}+{{\kappa }_{ss}}{{R}^{+}} & 0 & 0  \\
		-\rho_{d} T_{{}}^{-}T_{{}}^{+}{{\alpha }_{{}}} & 0 & -{{\kappa }_{s}}{{R}^{+}}-{{\kappa }_{ss}}{{R}^{+}} & 0  \\
		0 & 0 & 0 & -{{\kappa }_{s}}{{R}^{+}}-{{\kappa }_{ss}}{{R}^{+}}  \\
	\end{matrix} \right].
	$}}
\end{equation}

Since the light source has a linear polarizer at 0$^\circ$, the incident light is linearly polarized so its Stokes vector is ${{\mathbf{s}}_{i}} = [1, 1, 0, 0]^{\top }$. Plugging this into Equation~\eqref{eq:light_transport} leads to the reflected Stokes vector~${{\mathbf{s}}_{o}}$ as
\begin{equation}
	\label{eq:horizontal_polar_light}
	{{\mathbf{s}}_{o}}=S
	\mathbf{P}\mathbf{s}_i=S
	\left[ \begin{matrix}
		\rho_{d} T_{{}}^{+}T_{{}}^{+}-\rho_{d} T_{{}}^{-}T_{{}}^{+}\beta +{{\kappa }_{s}}{{R}^{+}}+{{\kappa }_{ss}}{{R}^{+}}  \\
		-\rho_{d} T_{{}}^{-}T_{{}}^{+}\beta +{{\kappa }_{s}}{{R}^{+}}+{{\kappa }_{ss}}{{R}^{+}}  \\
		-\rho_{d} T_{{}}^{-}T_{{}}^{+}\alpha   \\
		0  \\
	\end{matrix} \right].
\end{equation}



Given the fixed polarization filters, our camera captures a four-channel color image including four different polarization orientations. The Stokes image $\mathbf{I}$ can then be expressed as
\begin{align}
	\label{eq:polar_camera}
	\mathbf{I} & = 
	\left[ \begin{matrix}
		{{I}_{0}}  \\
		{{I}_{90}}  \\
		{{I}_{45}}  \\
		{{I}_{135}}  \\
	\end{matrix} \right]= \left[ \begin{matrix}
	1 & 1 & 0 & 0  \\
	1 & -1 & 0 & 0  \\
	1 & 0 & 1 & 0  \\
	1 & 0 & -1 & 0  \\
\end{matrix} \right] \mathbf{s}_{o}
 \\ \nonumber
	&=\frac{S}{2} 
	\resizebox{0.7\linewidth}{!}{
	\mbox{\fontsize{10}{12}\selectfont $
	\left[ \begin{matrix}
		\rho_{d} T_{{}}^{+}T_{{}}^{+}-2\rho_{d} T_{{}}^{+}T_{{}}^{-}\beta +2{{\kappa }_{s}}{{R}^{+}}+2{{\kappa }_{ss}}{{R}^{+}}  \\
		\rho_{d} T_{{}}^{+}T_{{}}^{+}  \\
		\rho_{d} T_{{}}^{+}T_{{}}^{+}-\rho_{d} T_{{}}^{+}T_{{}}^{-}\beta +{{\kappa }_{s}}{{R}^{+}}+{{\kappa }_{ss}}{{R}^{+}}-\rho_{d} T_{{}}^{-}T_{{}}^{+}\alpha   \\
		\rho_{d} T_{{}}^{+}T_{{}}^{+}-\rho_{d} T_{{}}^{+}T_{{}}^{-}\beta +{{\kappa }_{s}}{{R}^{+}}+{{\kappa }_{ss}}{{R}^{+}}+\rho_{d} T_{{}}^{-}T_{{}}^{+}\alpha   \\
	\end{matrix} \right]
	$}}.
\end{align}

While previous work required a dense set of angular samples~\cite{Baek2018}, the four components of $\mathbf{I}$ provide valuable enough information about the pBRDF. First, $I_{90}$ refers to the diffuse lobe, so we define our first \emph{diffuse shading} observation ${I}^{d}$ as 
\begin{align}
	\label{eq:polar_diffuse_term}
	{{I}^{d}}&=S \rho_{d} T^{+}T^{+} =2{{I}_{90}}.
\end{align}

Information about the diffuse term can also be obtained from the subtraction of $I_{45}$ and $I_{135}$, which we define as a \emph{diffuse polarization} observation ${I}^{\alpha }$ as
\begin{align}
	\label{eq:polar_diffuse_polarization}
	{{I}^{\alpha }}&= S \rho_{d} T_{{}}^{-}T_{{}}^{+}\alpha  ={{I}_{135}}-{{I}_{45}}.
\end{align}

Last, by substituting $I_{90}$ in $I_{0}$, we can obtain a combination of specular reflection, single scattering and oriented diffuse parameters. We define this combination as the \emph{specular-dominant polarization} observation $I^s$ as
\begin{align}
	\label{eq:polar_specular_term}
	{{I}^{s}}&=S\left( {{\kappa}_{s}}{{R}^{+}}+{{\kappa }_{ss}}R^{+}-\rho_{d} T_{{}}^{-}T_{{}}^{+}{{\beta }_{{}}} \right)={{I}_{0}}-{{I}_{90}}.
\end{align}

In the following section, we show how to leverage these single-image observations to optimize spatially-varying pBRDF parameters and surface geometry from \textit{multiple} views.

\section{Multiview Reconstruction of Polarimetric SVBRDF and Shape}
\label{sec:multiview}

\subsection{Overview}
\label{sec:method-overview}
\begin{figure}[tpb]
\vspace{-2mm}%
  \centering%
  \footnotesize%
\includegraphics[width=1.0\linewidth]{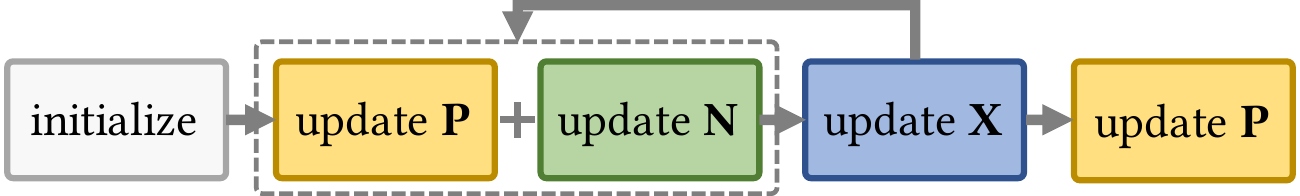}%
\vspace{-3mm}%
  \caption{\label{fig:workflow}%
Overview of our method. Iteratively, we alternate joint estimations of polarimetric SVBRDF $\mathbf{P}$ and shading normals $\mathbf{N}$ with reconstructions of 3D geometry $\mathbf{X}$.
}
  \vspace{-2mm}
\end{figure}

An overview of our method can be seen in Figure~\ref{fig:workflow}. Our input consists of a set of $K$ unstructured polarimetric photographs $\mathbf{I} = \{ I_{k}\}$ taken with our portable hardware (Section~\ref{sec:handheld-setup}) and interpreted through our image formation model (Section~\ref{sec:image-formation model}). First, an initialization step linearizes $I_{k}$ and obtains camera parameters and a rough base geometry using conventional 3D reconstruction techniques (Section~\ref{sec:initialization}). We then iteratively reconstruct polarimetric SVBRDF information $\mathbf{P}$ and shading normals $\mathbf{N}$ through inverse rendering (Section~\ref{sec:method-pbrdf}), and then reconstruct detailed 3D geometry $\mathbf{X}$ by means of Poisson surface reconstruction (Section~\ref{sec:method-geometry-opt}). In the last step, we update the polarimetric SVBRDF $\mathbf{P}$ from the final 3D geometry.

\subsection{Initialization}
\label{sec:initialization}

From our cross-polarized diffuse observations in the captured data, defined as $I^d$ in Equation~ \eqref{eq:polar_diffuse_term}, we first obtain a dense 3D point cloud with normal information and camera poses, which will provide the correspondences between 3D points and pixels in multiple views, using Structure-from-Motion (SfM)~\cite{schonberger2016pixelwise}. To mitigate the inherent reconstruction errors from SfM, we first create a low-resolution mesh ($2^{7}$-level octree in the voxel grid) using the screened Poisson surface reconstruction~\cite{kazhdan2013screened}, then subdivide it to obtain a finer mesh ($2^{9}$-level). This finer mesh is used as the initial geometry of our iterative technique, and might still miss geometric details due to the limitations of SfM. The geometry stage of our iterative algorithm (Section~\ref{sec:method-geometry-opt}) recovers these details.

\subsection{Optimizing Polarimetric SVBRDF and Normals}

\label{sec:method-pbrdf}
We jointly optimize polarimetric SVBRDF and normals by minimizing the sum of four loss terms as
\begin{equation}
	\label{eq:optimization}
	\underset{\eta,\sigma_{s},\rho_s,{\rho}_{ss},{\rho_d},\mathbf{n}}{\min}\left( {{\lambda }_{1}}{{\mathcal{L}}_{\psi}}+{{\lambda }_{2}}{{\mathcal{L}}_{d}}+{{\lambda }_{3}}{{\mathcal{L}}_{s}}+{{\lambda }_{4}}{{\mathcal{L}}_{\phi}} \right),
\end{equation}
where ${{\mathcal{L}}_{\psi}}$ is the refractive index loss,
${{\mathcal{L}}_{d}}$ is the diffuse loss,
${{\mathcal{L}}_{s}}$ is the specular and single scattering loss, and
${{\mathcal{L}}_{\phi}}$ is the normal loss. We set the 
weights to $\lambda_{\{1,3\}}=1$, $\lambda_{\{2,4\}}=100$.
For efficiency, ${{\mathcal{L}}_{d}}$ and ${{\mathcal{L}}_{s}}$ are calculated by linear optimization, while ${{\mathcal{L}}_{\psi}}$ and ${{\mathcal{L}}_{\phi}}$ are computed by nonlinear optimization, using the sequential quadratic programming method.

\paragraph{Refractive index loss ${{\mathcal{L}}_{\psi}}$} 
Existing approaches~\cite{Baek2018,Baek2021} require dense angular samples, which force fixed acquisition setups. Our novel refractive index loss overcomes such angular density requirements, providing a more flexible capture setup.
The key insight is that the degree of polarization (DoP) of the diffuse reflection depends on the refractive index and normal orientation can be represented as a ratio of Fresnel transmission functions ($\psi =|{T^{-}}/{T^{+}}|$). We thus define ${{\mathcal{L}}_{\psi}}$ as a loss function over DoP as
\begin{equation}
	\label{eq:dop_loss}
	{{\mathcal{L}}_{\psi}}=\sum\nolimits_{k=1}^{K}{{{w}^{v}_{k}}}{{\left( \hat{\psi }\left( \eta ,{{\theta }_{o,k}} \right)-\psi _{k}^{{}} \right)}^{2}},
\end{equation}
where $\hat{\psi}$ is the predicted DoP value for the $k$-th view, represented as $|{T^{-}}/{T^{+}}|$. Fresnel transmissions depend on normal orientation ${{\theta }_{o,k}}$ and index of refraction $\eta$ (both being optimized with this loss function). The visibility weight $w^{v}_{k}$ for this view is calculated as $w^{v}_{k}={{v}_{k}}/\sum\nolimits_{i=1}^{K}{{{v}_{i}}}$, where $v=\{0,1\}$.

Our input does not provide a direct observed DoP value $\psi_k$, so we approximate it from our observations $I^d$, $I^\alpha$, and $I^s$ (Equations~\eqref{eq:polar_diffuse_term},~\eqref{eq:polar_diffuse_polarization}, and~\eqref{eq:polar_specular_term} in Section~\ref{sec:image-formation model}). Our diffuse Mueller matrix, described in Equation~\eqref{eq:diffuse_mueller}, depends on two parameters: $\alpha=\sin (2\phi)$, and $\beta=\cos (2\phi)$. This leads to two diffusely polarized images $I^\alpha=-{S}{{\rho_d }}T^{-}T^{+}{{\alpha }}$ and  $I^\beta=-{S}{{\rho_d }}T^{-}T^{+}{{\beta }}$, from which diffuse polarization can be obtained as $\Gamma = \sqrt{ ({{I}^{\alpha}} )^{2}+ ( {{I}^{\beta}} )^{2}}$, and  $\psi_k$ becomes

\begin{equation}
	\label{eq:dop_specular}
	{{\psi }}=\left|\frac{\Gamma}{I^d}\right|	=\left|\frac{S \rho_{d} T^{+}T^{-}}{S \rho_{d} T^{+}T^{+}}\right|,
\end{equation}
where $\Gamma$ and $I^d$ have their three color channels averaged into a single value. For simplicity, note that we have removed the subindex~$k$ referring to each individual view.

Different from $I^\alpha$ and $I^d$,  ${{I}^{\beta}}$ cannot be directly obtained from our input images. Instead, we obtain ${{I}^{\beta}}$ by subtracting the polarimetric specular components ${{\kappa }_{s}}SR^{+}$ and ${{\kappa }_{ss}}SR^{+}$ from our specular-dominant polarization observation $I^s$, as
\begin{equation}
	\label{eq:beta_term}
	{{I}^{\beta}}={{I}^{s}}-{{\kappa }_{s}}SR^{+} - {{\kappa }_{ss}}SR^{+},
\end{equation}
where the parameters' values come from the previous iteration. For the first iteration, we approximate $I^\beta \approx I^s$.  


\paragraph{Diffuse Loss ${{\mathcal{L}}_{d}}$} 
We formulate our diffuse loss ${{\mathcal{L}}_{d}}$ by comparing the predicted diffuse image $\hat I^d_k$ with the captured image $I^d_k$ for the $k$-th view as 
\begin{equation}
	\label{eq:diffuse_loss}
	{{\mathcal{L}}_{d}}={\sum\nolimits_{k=1}^{K}{{{w}^{v}_{k}}{{\left({{\hat I}_{k}^{d}} -{{I}_{k}^{d}} \right)}^{2}}}},
\end{equation}
where the Fresnel transmission in ${\hat I}_{k}^{d}$, shown in Equation~\eqref{eq:polar_diffuse_term}, is computed from the refractive index of the previous iteration (initially set to $\eta=1.5$). Losses of each color channel are summed together.

\paragraph{Specular and single scattering loss ${{\mathcal{L}}_{s}}$} 
%
Despite using multiple photographs as input, the number of samples is often insufficient to estimate \emph{per-vertex} specular and single scattering parameters, especially for very narrow specular lobes. To solve this, the approach by existing reconstruction methods~\cite{Nam2018,Baek2018, 
wu2015appfusion,MicroscaleSVBRDF:SIGA:2016,Alldrin2008,lawrence2006inverse,chen2014reflectance,zhou2016sparse}
is to obtain the specular parameters of vertices with no suitable view by clustering vertices and assigning them the same specular parameters, which is prone to appearance parameter estimation errors due to imperfectness in the clustering algorithm. \NEW{We instead apply a novel \emph{specular augmentation} strategy  to increment the number of per-vertex observations, by generating virtual samples from the specular and single scattering parameters of its local neighborhood.}



Then, our specular and single scattering loss term is composed of two parts. The first, as expected, is the difference between the predicted specular and single scattering polarization $\hat I_k^s$ and the captured $I_k^s$ for the $k$-th view, and the second is the difference between the predicted specular and single scattering $\hat I_m^s$ and the sample $\tilde I_m^s$ for a \textit{virtual} $m$-th \NEW{observation}, as
\begin{equation}
	\label{eq:specular_loss}
	{{\mathcal{L}}_{s}}
	=\sum\nolimits_{k=1}^{K} {w}^{v}_{k} \left( {\hat I}^{s}_k - {I}^{s}_k \right)^2+\lambda_g {\sum\nolimits_{m=1}^{M}{{w}^{a}_{m}{{\left( {{{\hat{I}}}^{s}_{m}}-{\tilde{I}^{s}_{m}} \right)}^{2}}}},
\end{equation}
where 
${w}^{a}$ is a normalized $\cos ({{\theta_h })}$ weight, 
$m \in [1,...,M]$ is the \NEW{virtual observation} index
(we set $M$ to 180 to cover 90$^\circ$),
$\lambda_g$ is the \NEW{specular augmentation} loss weight, set to $0.1$ so that the loss function prioritices the match from real views over the virtual \NEW{observations}. Losses for each color channel are summed together.


To generate these new virtual views $\tilde I_m^s$, we assume that many similar vertices share similar material features, and thus vertices that share similar refractive index and diffuse albedo are more likely to share also specular and single scattering properties.
We therefore define a four-dimensional feature vector that includes the refractive index and all three channels of the diffuse albedo, and use $K$-means to cluster vertices according to such features. 

%
We then regress the specular and single scattering model parameters for each cluster by minimizing
\begin{equation}
	\label{eq:pbrdf_optimization_cluster}
	\underset{\eta,\sigma_{s},\sigma_{ss},\rho_s, \rho_{ss}, \Delta\theta_h}{\min}\left( {{\mathcal{L}}_{\psi}}+{{\mathcal{L}}_{s}'} + \mathcal{L}_\theta \right),
\end{equation}
where ${{\mathcal{L}}_{s}'}$ represents the first loss term only in Equation~\eqref{eq:specular_loss},
$\Delta\theta_h$~is a variable that accounts for potential $\theta_h$ errors in the measured samples (to which specular parameters are specially sensitive),
and $\mathcal{L}_{\theta} = (\Delta\theta_h)^2$ is a regularization loss for $\Delta\theta_h$. This yields specular parameters per cluster. In contrast to previous works~\cite{Nam2018,Baek2018}, the optimized specular parameters are not directly set for the vertices (which is very prone to errors over edges and high-frequency textures) but used to generate novel views to include in the loss function. We render these new views as per-vertex $\tilde I^s$ observations distributed uniformly in $\theta_h$ from zero to 90$^\circ$ at 0.5$^\circ$ intervals and include them into the loss function as defined in Equation~\eqref{eq:specular_loss}.

Last, the calculation of the loss function for the real views requires the evaluation of the specular-dominant observations $I^s$ as in Equation~\eqref{eq:polar_specular_term}.  $I^s$ includes a  term $-\rho_{d} ST_{{}}^{-}T_{{}}^{+}{{\beta }_{{}}}$ which is noisy, relatively weak and cancels out at different views, so we approximate it to zero. We assume dielectric surfaces, and thus we use a single channel for specular albedo and three channels for the single scattering albedo.


\paragraph{Normal loss ${{\mathcal{L}}_{\phi}}$} 
\label{sec:opt-normals}
The zenith angle $\theta_o$ of the normal is already accounted for by the aforementioned loss terms for the index of refractive, diffuse, and specular polarization, since it affects the Fresnel terms. Therefore, our normal loss term deals only with the azimuth angle $\phi_o$. However, $\phi_o$ contains an ambiguity of $\pi$ radians, which leads to errors in the reconstructions; this is known as the azimuthal ambiguity \cite{Atkinson2006,kadambi2015polar}.
In our work, we minimize the errors caused by this ambiguity leveraging our multiple observations from different view angles. Our normal loss is formulated as
\begin{equation}
	\label{eq:azimuth_loss}
	{{\mathcal{L}}_{\phi}}=\sum\nolimits_{k=1}^{K}{2w_{k}^{p} \left( 1 - \cos \left( 2{{\hat \phi }_{o,k}}-2{{\phi }_{I,k}} \right) \right)},
\end{equation}
where $\hat \phi_{o,k}$ is the azimuth angle of the estimated normal at the $k$-th view and
$\phi_{I,k}$ is the observed azimuth angle from diffuse polarization computed by ${2{\phi }_{I,k}}={{\tan }^{-1}}( {{{I}_{k}^{\alpha}}}/{{{I}_{k}^{\beta}}} )$.
The diffuse polarization weight 
$w_{k}^{p}=w^v_{k}\Gamma^2_k$ 
prioritizes stronger diffuse polarization signals.

\subsection{Optimizing Geometry}
\label{sec:method-geometry-opt}
After estimating the polarimetric SVBRDF $\mathbf{P}$ and shading normals~$\mathbf{N}$, we then reconstruct the geometry~$\mathbf{X}$ so that it agrees with the polarimetric observations. 
We update this geometry with estimated shading normals, following a recent work \cite{Nam2018}.
Since the shading normals $\mathbf{N}$ may contain high-frequency noise, we choose the screened Poisson method \cite{kazhdan2013screened}, designed to reconstruct implicit surfaces in a voxel grid in a coarse-to-fine approach. 
This leads to a robust performance when integrating noisy surface normals into 3D geometry.

In the last iteration, after the denoised final shading normals~$\mathbf{N}$ are obtained from the final geometry~$\mathbf{X}$, we only optimize the polarimetric SVBRDF $\mathbf{P}$ one last time.

\section{Implementation Details}
\label{sec:implementation}
\paragraph{Experimental setup.}
%
We build our capture setup from two off-the-shelf components: a polarization machine vision camera (LUCID PHX050S-QC), and a flashlight (Nikon Speedlight SB24) covered with a linear polarization film.
The two components are supported together with a  custom 3D-printed structure (see Figures~\ref{fig:teaser} and~\ref{fig:coordinates}b). 
To properly capture HDR specular reflections we rely on multi-bracketing~\cite{Debevec:1997:RHD}, using three flash levels at \{$1/4$,$1/8$,$1/16$\} of its maximum intensity per view. We capture around $100$ and $300$ views per object. The exact number of views for each of our results can be found in the supplemental material.



\paragraph{Calibration}
To estimate the exact geometric relation between the camera and the flash (since they cannot be perfectly coaxial), we use the mirror reflections of the flashlight on multiple stainless spheres with a known radius. We place them on a grid at regular $25$mm intervals, and estimate the light position with respect to the camera by accounting for the trigonometry of the specular reflections, following \citet{Lensch:2003:IMAGE}. 
\NEW{In our device, the separation between the camera and the light source is 5\,cm, so the angle between light and camera rays is $\sim$3.5$^\circ$
when the captured object is at a distance of around 80--100\,cm.}
The intrinsic geometry of the polarization camera is calibrated using the method by \citet{zhang_calibration:2000}. Last, we also calibrate color reproduction by estimating a linear matrix from RAW RGB channels to the sRGB color channels using a Macbeth ColorChecker. Note that these calibrations only need to be performed once per device.

\paragraph{Optimization} 
The size of our reconstructed models varies between $500,000$ and $800,000$ vertices, with pBRDF parameters stored as per-vertex attributes. We optimize Equation~\eqref{eq:optimization}  by means of a nonlinear optimizer, sequential quadratic programming (SQP), using \texttt{fmincon} in MATLAB. One iteration takes about one hour in a desktop computer with an Intel i9-12900 CPU 3.2 GHz and 64 GB of memory
and an NVIDIA Titan RTX GPU (similar to previous approaches). Final results are produced after ten iterations.



\section{Validation}
\label{sec:validation}

\subsection{pBRDFs Models}
\label{sec:practical-model-validation}
\begin{figure*}[tpb]
	\centering	\includegraphics[width=\linewidth]{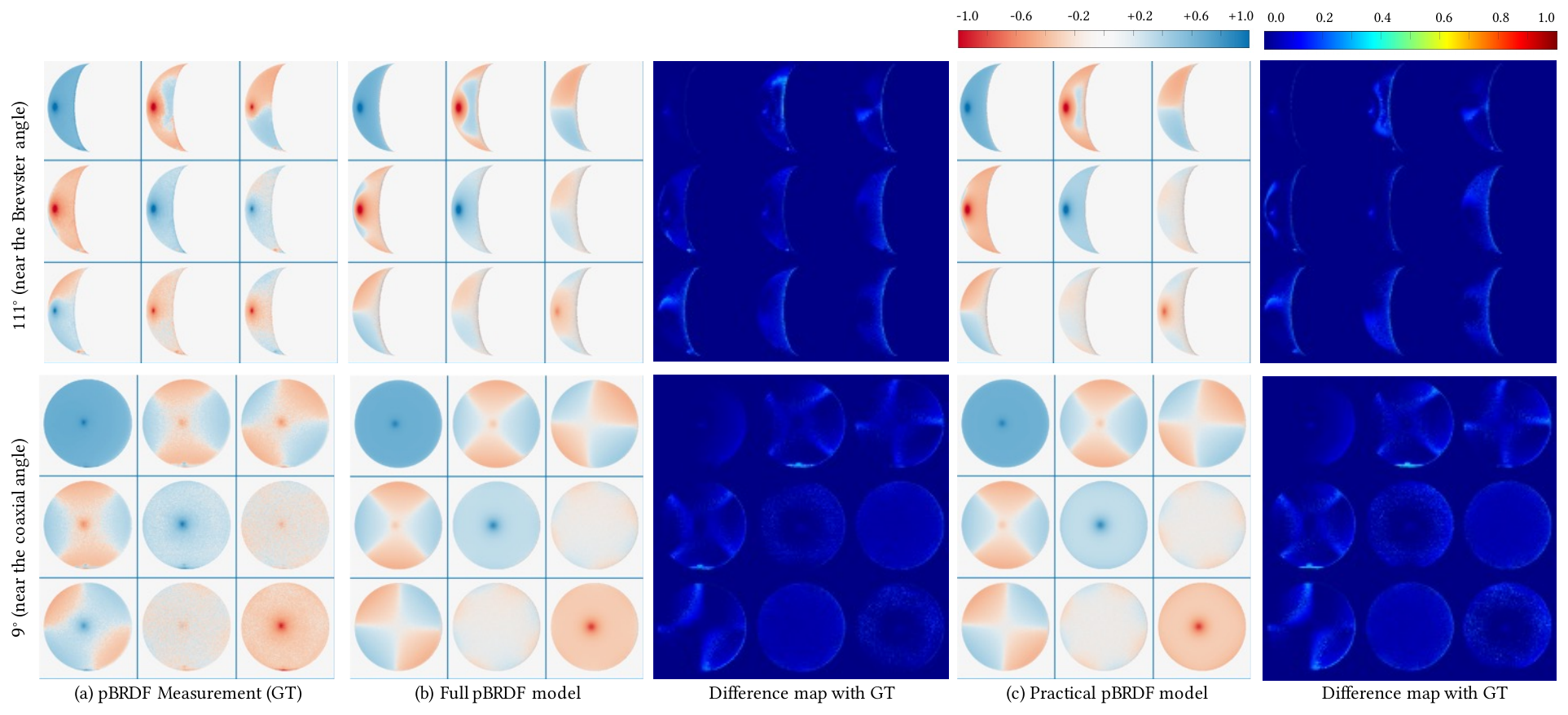}%
	\vspace{-5mm}%
	\caption[]{\label{fig:validation-simple}%
	Validation of our pBRDF models by comparing the resulting Mueller matrices, at two different angles. (a) Ground-truth, captured data of an organic thermoplastic polymer object (PEEK) \cite{Baek2020}.  (b) Our full pBRDF model and its difference map with respect to the ground truth. (c) Our practical model and its difference map. Despite the simplifications introduced in our practical model, it offers accurate results while being suitable for optimization.  
	}%
\end{figure*}

First, we evaluate the accuracy of our proposed physically-based and practical pBRDF models against actual pBRDF measurements by \citet{Baek2020}, both at 111$^\circ$ near the Brewster angle of the object, and at 9$^\circ$, closest to our coaxial setup. Figure~\ref{fig:validation-simple} shows the resulting Mueller matrices, along with false-color difference maps. Despite its simplifications, our practical model offers accurate results while being suitable for optimization.

\subsection{Single Scattering}
\label{sec:result-scattering}

\begin{figure}[tpb]
	\centering	\includegraphics[width=\columnwidth]{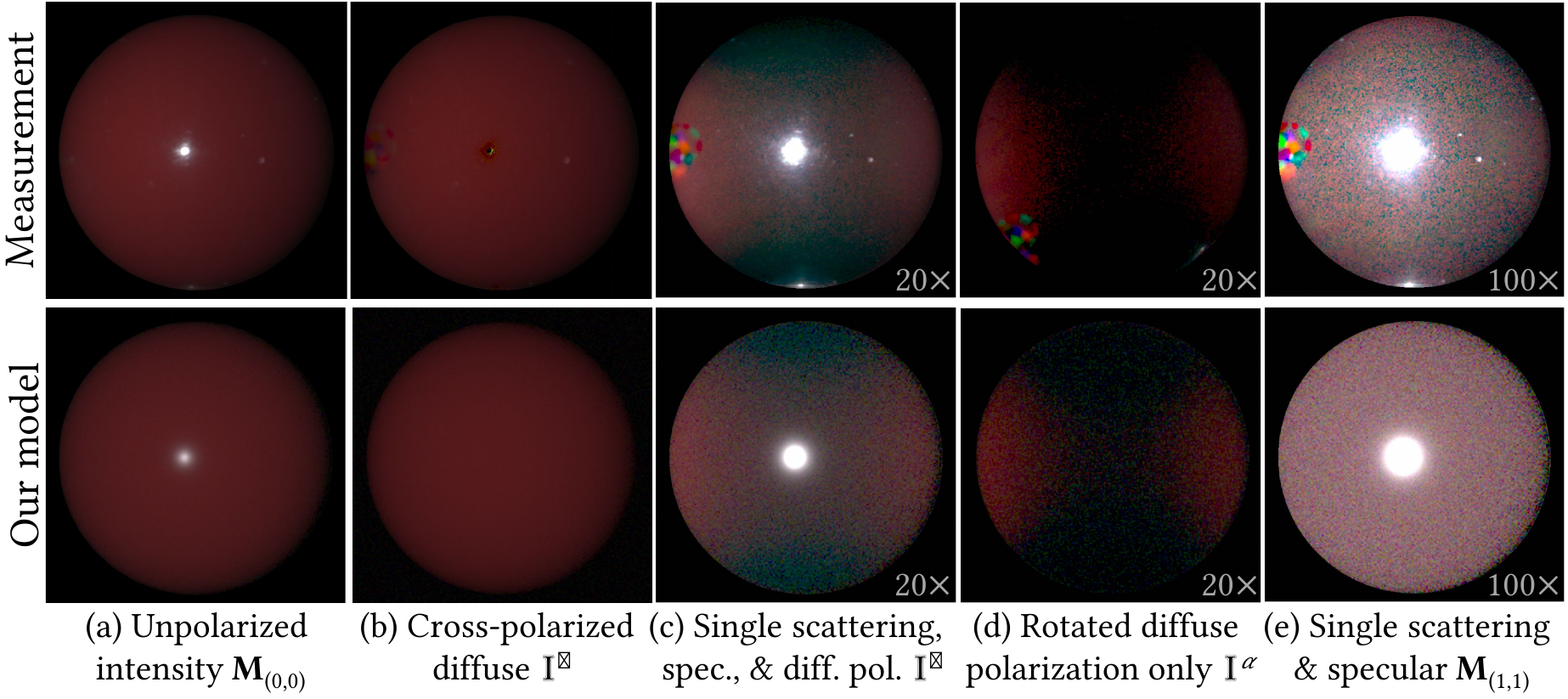}%
	\vspace{-3mm}%
	\caption[]{\label{fig:red_billiard}%
	Top: measured polarimetric data from a red billiard ball~\cite{Baek2020} (the rainbow speckles are caused by lens flare effects in the capture system). Bottom: results from our model. Columns (c), (d), and (e) have been amplified by the indicated factors for visualization purposes.

	}%
\end{figure}

\NEW{Our new single scattering component, as defined by Equation~\eqref{eq:single_scattering_transport2}, is different from the specular component in two ways: first, it presents a \textit{colored} albedo which is different from the diffuse component, while the specular reflection retains the color of the illumination. Second, it has an independent roughness parameter, which approximates two Fresnel transmissions and a single subsurface interaction with an unknown phase function. The combination of colored albedo and independent roughness leads to clearly different single scattering and specular components in most scenes. Figure~\ref{fig:red_billiard} shows side-by-side comparisons between captured polarimetric data from a red billiard ball ~\cite{Baek2020} and the results optimizing our practical pBRDF model (Section~\ref{sec:image-formation model}). We use the measurements closest to the coaxial angle (9$^\circ$ between the light and the sensor in~\cite{Baek2020}, compared to  3.5$^\circ$ in our setup). This shows a good agreement with measured data.}

\subsection{Index of Refraction}
\label{sec:result-ri}

The index of refraction is the most unstable parameter when optimizing our model, since it affects all the lobes of the polarimetric SVBRDF in different ways, with ambiguities regarding how its variation affects appearance. We validate the accuracy of our optimized index of refraction against real-world materials from the existing pBRDF dataset of~\citet{Baek2020}. 
As input, we use synthetic views generated from the captured dataset. 
The ground truth index of refraction from measured data is obtained from the Brewster angle as explained in the comprehensive work of \citet{collett2005field}. 

Table~\ref{tab:refractive_index} shows the results. Despite needing only four linear polarization measurements per view instead of the dense measurements of full ellipsometry, the accuracy of our method is consistently high.

\begin{table}[htpb]
	\caption[]{\label{tab:refractive_index}%
Comparison between the ground truth index of refraction of several objects measured by \citet{Baek2020} and our optimized results.
}
\vspace{-3mm}
	\begin{tabular}{l c c c c c}
		\thickhline
		Material & $\eta_{m}$  & $\eta_{ours}$ & error\\
		\hline
		White billiard & 1.463 & 1.465 & 0.10\%\\
		Red billiard & 1.485  & 1.476 & 0.61\%\\
		Green billiard & 1.503 & 1.476 & 1.80\%\\
		POM & 1.462 & 1.457 & 0.34\%\\		
		Fake pearl & 2.295 & 2.244 & 2.22\%\\
		Yellow silicone & 1.303  & 1.337 & 2.61\%\\
		PEEK & 1.663 & 1.617 & 2.77\%\\
		 \hline
		Average & &  & 1.49\% \\
		\thickhline
	\end{tabular}
\vspace{-5mm}
\end{table}

\subsection{Impact of Specular Augmentation}
\label{sec:iteration}
\begin{figure}[tpb]
	\centering	
	\includegraphics[width=\linewidth]{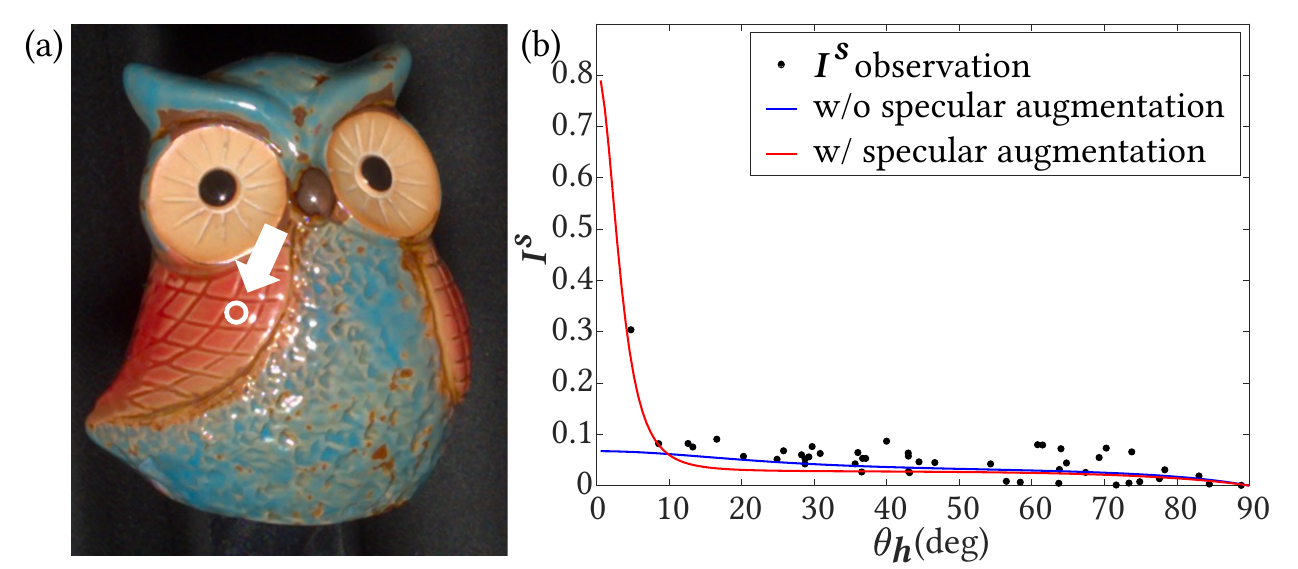}%
	\vspace{-3mm}%
	\caption[]{\label{fig:specular_augmetation}%
	The sparse nature of our input images would lead to wrong estimations of specular reflections. (a) Example input image without specular information in the highlighted area. (b) The resulting specular reflection functions with and without data augmentation (red and blue lines, respectively).  
	}%
	\vspace{-3mm}
\end{figure}

Next, we evaluate the impact of our specular augmentation algorithm (Section~\ref{sec:method-pbrdf}).
Figure~\ref{fig:specular_augmetation}(a) shows one of our input photographs; for this particular view, there are almost no specular reflections in the vertex indicated by the arrow. Figure~\ref{fig:specular_augmetation}(b) shows a scatter plot of the captured specular-dominant samples $I^s$, where each black dot indicates a specular observation projected onto the normal space of the vertex in the 3D object. The blue line shows the profile of the resulting regressed specular reflection, which would translate into an overly diffuse appearance. The red line represents the optimized profile after our specular data augmentation method, which recovers the missing specular information and better fits the sparse captured data.  \NEW{For this owl scene, we obtain a PSNR of 34.0 dB with our proposed data augmentation, which drops to 30.8 dB without it. }

\section{Results}
\label{sec:results}

\subsection{Comparisons with Other Methods}
\label{sec:comparisons}

\begin{figure*}[tpb]
	\centering	
\includegraphics[width=\linewidth]{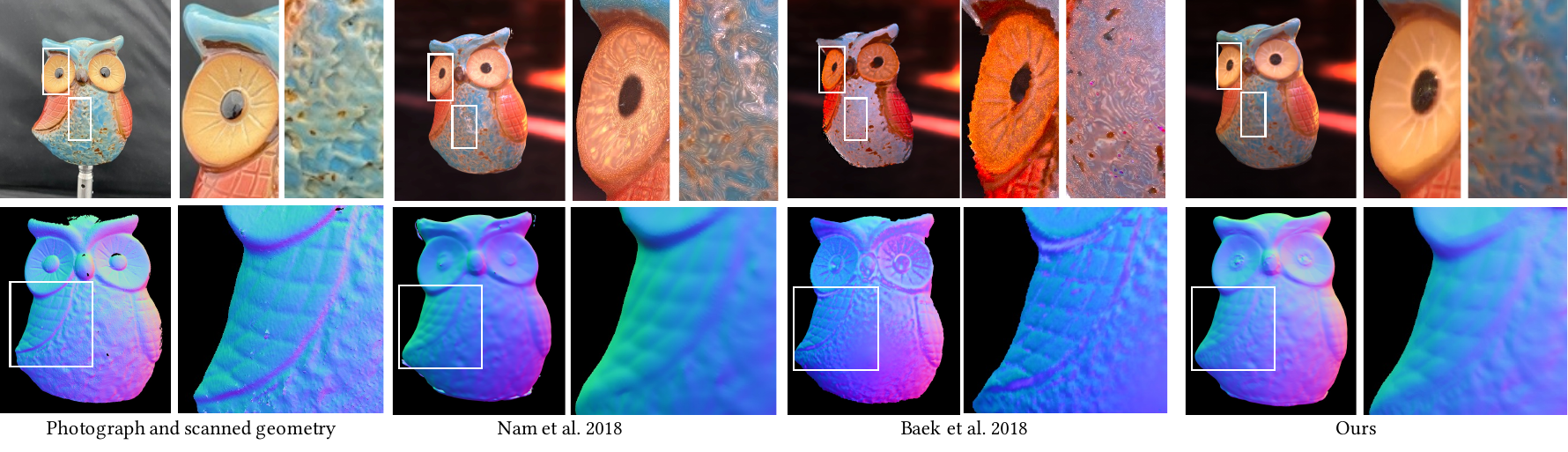}%
	\vspace{-3mm}%
	\caption[]{\label{fig:Env-rendering}%
Comparison of our method with two state-of-the-art techniques: a multiview 3D reconstruction method \cite{Nam2018}, and a polarimetric reconstruction method \cite{Baek2018}. The former yields visible overfitting errors in the reflectance, with overly smooth geometric details (see also Figure~\ref{fig:geometry_comparision}). The latter suffers from overfitting and clustering artifacts in reflectance, as well as geometric distortions in novel views.
 	}%
	\vspace{-3mm}
\end{figure*}

We first compare our shape and appearance reconstructions against two recent state-of-the-art methods. 
\citet{Nam2018} reconstruct 3D models with conventional SVBRDF from multiview stereo as input, while 
\citet{Baek2018} estimate polarimetric SVBRDF and surface normals from a given geometry in a single view. 
Figure~\ref{fig:Env-rendering} shows the results when generating novel views under different illumination. 
Nam's method estimates the normal distribution function as a tabulated function, which is convenient for representing specular reflection but it often causes visible artifacts in the recovered reflectance, due to overfitting during optimization. This is clearly visible in the eye region and, to a lesser extent, in the swirling patterns on the body. Baek's method, on the other hand, also suffers from overfitting artifacts in the reflectance, while showing additional clustering artifacts as well. Moreover, being a single-view method, it cannot reconstruct the full 3D geometry, which leads to visible distortions when rendering novel views.
In contrast, our method does not suffer from overfitting nor clustering artifacts in the reflectance, while our geometric reconstruction is more detailed, and closer to the ground truth from a commercial 3D laser scanner (NextEngine).

\begin{figure}[tpb]
	\centering	
\includegraphics[width=1.0\linewidth]{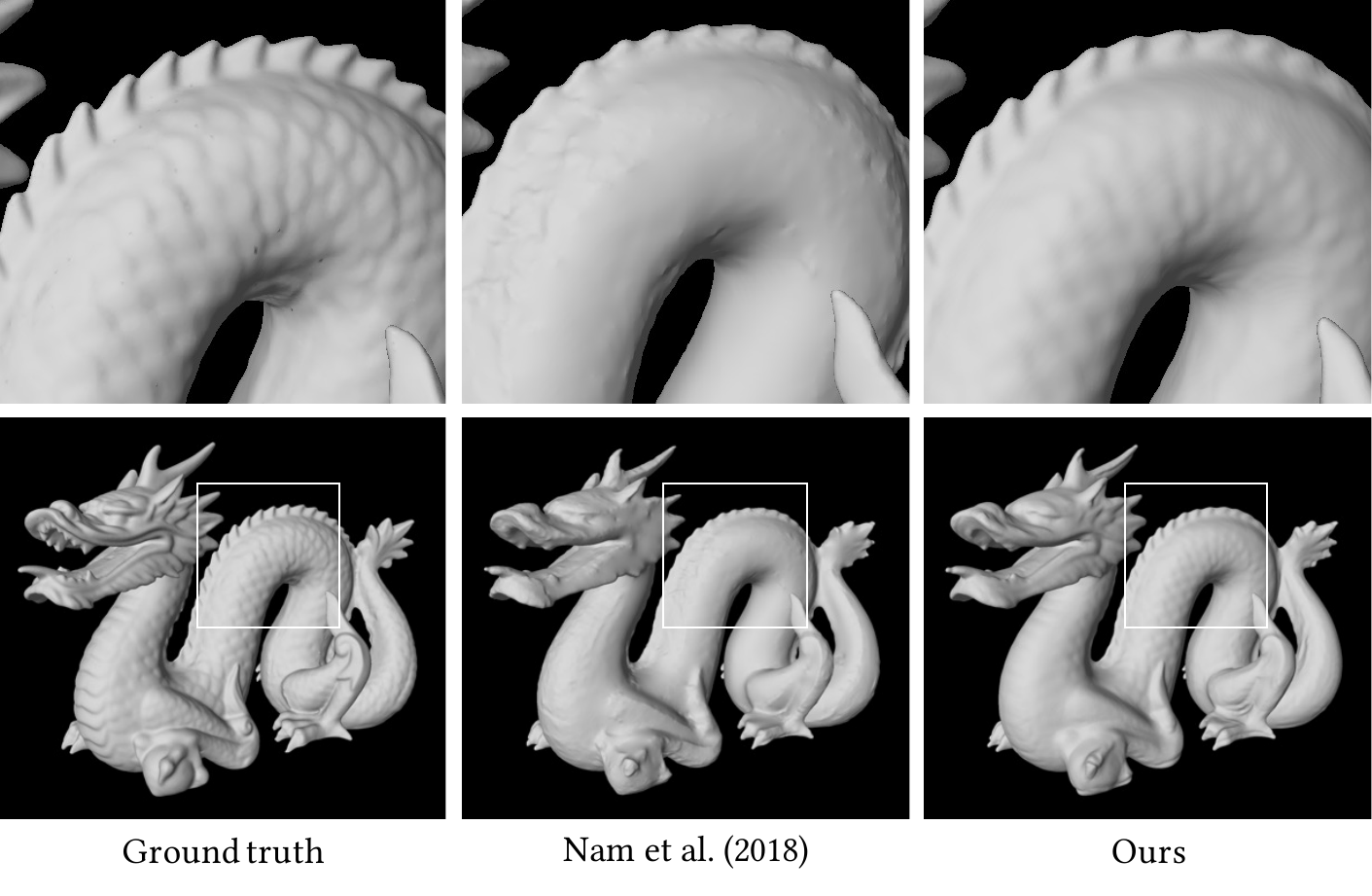}%
	\vspace{-3mm}%
	\caption[]{\label{fig:geometry_comparision}%
	Accuracy  of our 3D reconstructions against a state-of-the-art multiview reconstruction method~\cite{Nam2018}. Leveraging information about the polarization of light allows us to recover more geometrical details. 
	}%
	\vspace{-3mm}
\end{figure}

We further compare the accuracy of our resulting 3D reconstructions against the state-of-the-art multiview technique of~\citet{Nam2018}. As shown in Figure~\ref{fig:geometry_comparision}, our method recovers more detailed geometry, since it leverages polarization information at different angles, while Nam's work only takes into account variations in the intensity of light.

\subsection{Real-World Objects}
\label{sec:real-wold-objects}
\begin{figure*}[]
	\centering	\includegraphics[width=1.0\linewidth]{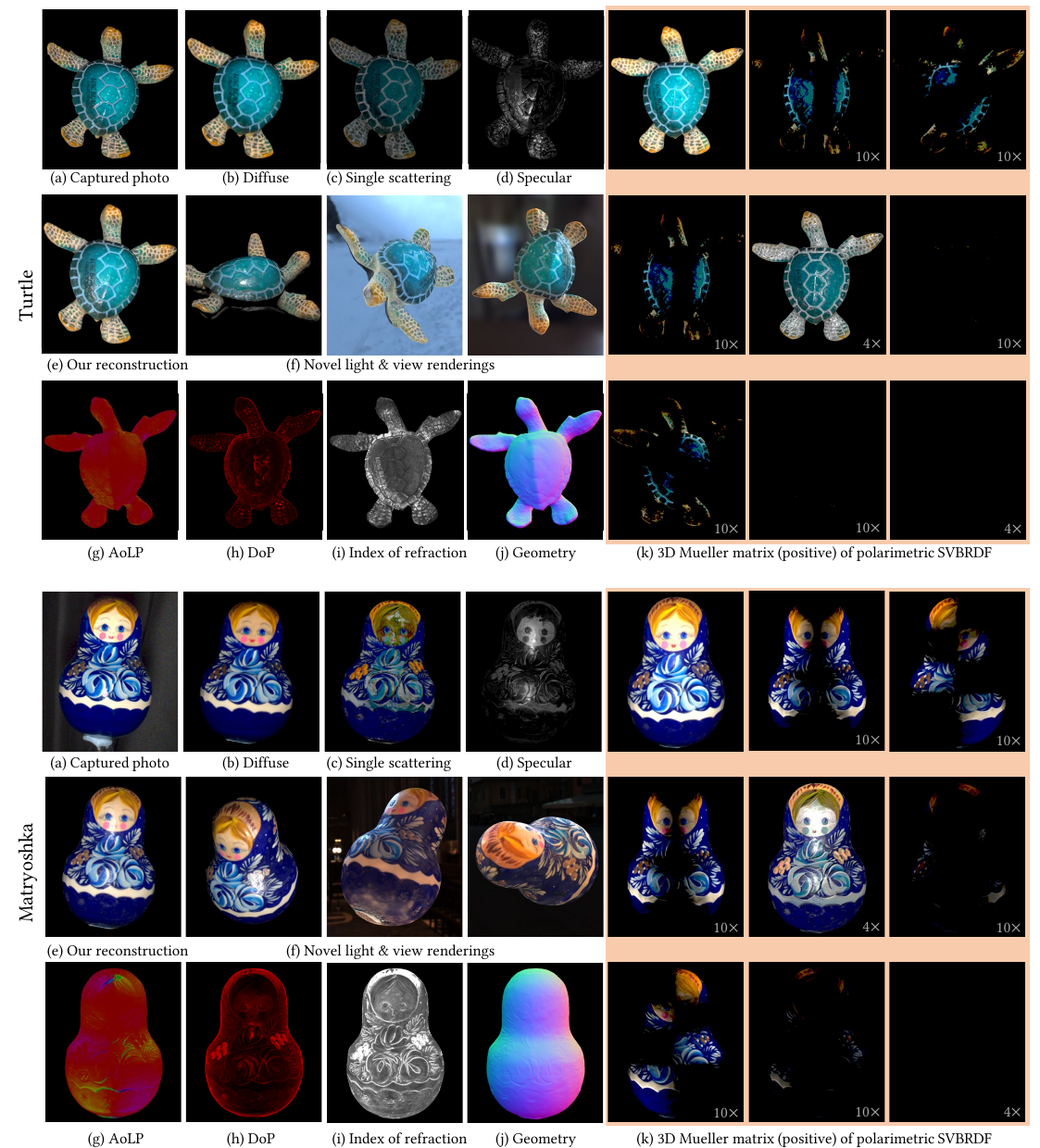}%
	\vspace{-3mm}%
	\caption[]{\label{fig:real_output}%
		Results for two examples of real objects (refer to the supplemental material for other objects and videos).
		(a) Captured photograph.
		(b) Diffuse albedo.
		(c) Single scattering.
		(d) Specular reflections.
		(e) Our reconstruction. 
		(f) Novel light and view renderings.
		(g) Angle of linear polarization image under linearly polarized light.
		(h) Degree of polarization.
		(i) Index of refraction.
		(j) 3D geometry including normal information.
		(k) 3D Mueller matrix (positive) for a single view. Note that the non-diagonal components have been multiplied by 10, while the $\mathbf{M}_{\{1,1\}}$ and $\mathbf{M}_{\{2,2\}}$ components have been multiplied by 4 for visualization purposes.
	}%
	\vspace{-3mm}
\end{figure*}

We have captured geometry and polarimetric SVBRDF of several real-world objects with our device and our sparse ellipsometry approach. Figure~\ref{fig:real_output} presents results for two objects; more results with different objects are available in our supplemental material, along with videos showing different poses and dynamic illumination. The angle of linear polarization (AoLP) and the degree of polarization (DoP) are visualized as proposed by~\citet{wilkie2010standardised}. The negative components of the 3D Mueller matrices for all objects are also included in the supplemental material.

\subsection{Application: Material Identification}
\label{sec:matrial-identification}
\begin{figure}[tpb]
	\centering	
	\includegraphics[width=1.05\columnwidth]{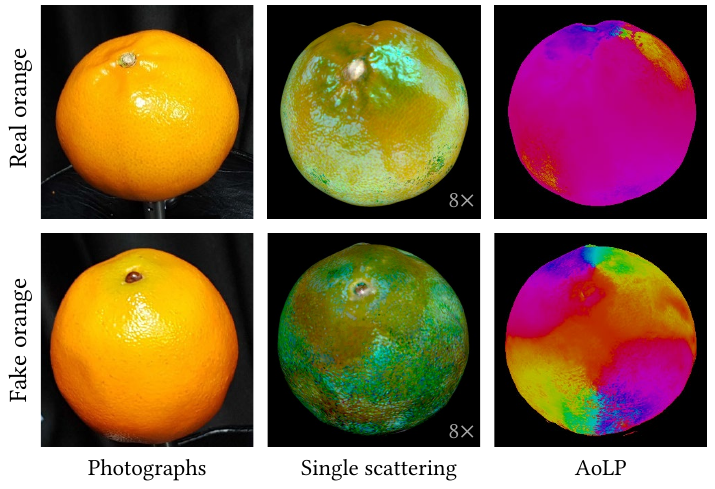}%
	\vspace{-3mm}%
	\caption[]{\label{fig:orange_comparison_d}%
Comparison of polarimetric information for a real (top) and a fake orange (bottom). Despite their similar appearance, polarimetric information allows us to distinguish between the two: the real orange presents a stronger single scattering component while preserving the polarization of incident light (as shown by the angle of linear polarization, AoLP). }%
	\vspace{-3mm}
\end{figure}

Polarimetric appearance information may be used to identify or discriminate between materials. As an illustrative example, Figure~\ref{fig:orange_comparison_d} shows the results of using our sparse ellipsometry technique on a real orange and a fake one made of plastic. The first column shows the two objects, whose appearance looks almost identical. The second column shows how the single scattering component of the real fruit is significantly higher than the plastic fake, which is hollow inside. The third column shows the angle of linear polarization under linearly polarized light (0$^\circ$). The strong single scattering in the real orange preserves the polarization of the incident light, which is not the case for the fake one.

\section{Discussion}
\label{sec:discussion}

\paragraph{Circular polarization} Our handheld setup captures four channels of linear polarization (0$^\circ$, 45$^\circ$, 90$^\circ$, and 135$^\circ$) and illuminates with a flashlight with fixed orientation (0$^\circ$). Therefore, the effect of circular polarization ($s_3$ of the Stokes vector) is not taken into account, which constitutes the main limitation of our method. Such circular polarization could be included by installing a retarder in front of the camera, but at the cost of losing one of the linear components (swapping $s_2$ and $s_3$, for instance). Another possibility would be to install and rotate optical components, which is something we avoid by design to keep our device truly portable and handheld.  Furthermore, retarders have a different effect per wavelength, so using them would require additional color calibration, making our approach less practical. 

Since circular polarization is omitted, our technique cannot estimate the appearance of metallic surfaces nor multiple scattering effects accurately. Nevertheless, geometry information strongly depends on linear polarization; dropping some of its components to add additional circular polarization information would lead to worse geometrical reconstructions, which would in turn lead to poor reflectance estimations. Furthermore, our polarimetric SVBRDF model is based on dielectric surfaces including Fresnel reflectance and transmittance, for which linear polarization also plays a significant role. \NEW{As observed in the measured pBRDF dataset of Baek et al.~\shortcite{Baek2020}, the amount of circular polarization is very small in a near-coaxial setup, for a dielectric surface illuminated with linear polarization. In any case, we expect our method to degrade gracefully with increasing circular polarization.}
%


\paragraph{Polarimetric SVBRDF model}
An interesting finding of our optimization procedure is that the value of the single scattering roughness $\sigma_{ss}$ consistently converges to a high value ($>0.9$). This is to be expected in our model since  $\sigma_{ss}$ represents the accumulated roughness of two rough Fresnel transmissions and a phase function, so it stands to reason that it behaves as an almost diffuse lobe with specular-like orientation and polarimetry. Leveraging this, $\sigma_{ss}$  could potentially be fixed to a high value during inverse rendering, saving up to an estimated $56\%$ in computation time with no significant accuracy penalty. This strategy, while reasonable, has not been applied to any result of this paper. \NEW{As a proof of concept for the owl scene, fixing $\sigma_{ss}=1$ leads to a PSNR of 33.9\,dB compared to 34.0\,dB using our full pipeline while fixing $\sigma_{ss}=0.9$ yields a PSNR of 34.2\,dB.}



\NEW{\paragraph{Impact of vertex resolution and initial geometry}
Vertex resolution is determined by the depth of the octree in the screened Poisson surface reconstruction method, which we set to $9$. This depth provides an adequate tradeoff between geometry resolution and convergence time. Our inverse rendering method optimizes per-vertex reflectance and geometrical properties (position and normals) with small-scale geometry shifts for mesoscale details. Similar to other state-of-the-art methods~\cite{Nam2018}, our results would degrade with poor initial geometry. In all our experiments, Poisson surface reconstruction has consistently provided a good enough initial geometry, but other methods (i.e., ball pivoting or visual hull) could be used if needed.}

\NEW{\paragraph{Failure cases} While robust, our method is not free of limitations. If the captured surface is extremely dark our DoP calculation will be affected by noise, which in turn degrades the accuracy of the refractive index. Also, in situations in which the specular and single scattering components cannot be easily disambiguated (e.g., a white rough surface or a very smooth surface with dominant specularity), insufficient observations will yield imperfect separations. This may lead to dark blotches, as the single scattering image in Figure~\ref{fig:real_output} (turtle scene, c) shows.}

\begin{table}[htpb]
\begin{center}
\caption{\label{tab:notations}
Notations used in this paper.}
\vspace{-3mm}
\resizebox{1.0\columnwidth}{!}{
\begin{tabular}{lll} \cline{1-3}
 & Symbol       & Description  \\ \cline{1-3}
\multirow{11}{*}{\rotatebox{90}{Mueller Mat. \& Stokes Vec.}}
& $\mathbf{M}$        & General Mueller matrix                                 \\
& $\mathbf{P}$        & pBRDF, a reflectance function of $\left(\bm{\omega}_{i},\bm{\omega}_{o}\right)$ \\
& $\mathbf{P}^{{d,s,ss}}$ & Diffuse/specular/single scattering pBRDF \\
& $\mathbf{C}_{{i} \to {n}}$ & Coordinate conversion Mueller matrix from light to plane of incidence   \\
& $\mathbf{C}_{{n} \to {o}}$ & Coordinate conversion from plane of incidence to camera system \\
& ${\mathbf{F}}^{{T}}_{{i,o}}$ & Incident/exitant Fresnel transmission matrix                               \\
& ${\mathbf{F}}^{{R}}$ &  Fresnel reflection matrix                          \\
& ${{\mathbf{D}}}$          & depolarization matrix                                 \\
& $\mathbf{s}$ & General Stokes vector consisting of four elements: $\left[s_0, s_1, s_2, s_3 \right]$  \\
& $\mathbf{s}_\text{i,o}$ & Stokes vectors of the light incident/exitant to an object surface \\
\cline{1-3}
\multirow{16}{*}{\rotatebox{90}{Geometry}}
& $\mathbf{y}_{{i,o}}$  &  $y$-axis of the light/camera coordinate system              \\
& $\mathbf{n}$          & Normal vector \\
& $\bm{\omega}_i$          & Incident light vector \\
& $\bm{\omega}_o$          & View vector\\
& $\mathbf{h}$       & Half way vector\\
& $\theta_{i,o}$          & Zenith angle between normals and the incident/exitant light                         \\
& $\theta_h$        & Zenith angle between normals and halfway vector  \\
& $\theta_d$          & Zenith angle between incident light and halfway vector                                  \\
& $\phi_{i,o}$          & Azimuth angle between the object plane of incidence \\
&           & and the $y$-axis of the incident/exitant \\
& $\varphi_{i,o}$        & Azimuth angle between the micro-facet plane of incidence \\
&  & and the $y$-axis of the incident/exitant light \\
& ${\alpha _{i,o}}$          & ${\alpha _{i,o}}=\sin (2{\phi _{i,o}})$     \\
& ${\beta _{i,o}}$          & ${\beta _{i,o}}=\cos (2{\phi _{i,o}})$     \\
& ${\chi _{i,o}}$          & ${\chi _{i,o}}=\sin (2{\varphi _{i,o}})$     \\
& ${\gamma _{i,o}}$          & ${\chi _{i,o}}=\cos (2{\varphi _{i,o}})$     \\
 \cline{1-3}
\multirow{7}{*}{\rotatebox{90}{Reflectance}}
& $\rho_d$          & Diffuse albedo                                              \\
& $\rho_s$          & Specular albedo \\
& $\rho_{ss}$          & Single scattering albedo \\
& $\sigma$  &  Surface roughness               \\
& $\eta$        & Refractive index \\
& $G$       & Smith's shadowing/masking function \\
& $D$       & GGX micro-facet distribution function \\
& $S$ & Light attenuation by shading, $S=\frac{\left( \mathbf{n}\cdot {{\mathbf{\omega }}_{i}} \right)}{d^2}$ \\
 \cline{1-3}
\multirow{12}{*}{\rotatebox{90}{Polarization-related variables}}
& $\psi$          & Degree of polarization                                                \\
& $\delta$       & Phase shift \\
& $T^{\perp,\parallel} _{{{i}}}$          & Fresnel incident transmission coefficients \\
& $T^{\perp,\parallel} _{{{o}}}$          & Fresnel exitant transmission coefficients \\
& $T^ + _{{{i}},{{o}}}$        & $({{T^ \bot _{{{i}},{{o}}} + T^\parallel _{{{i}},{{o}}}}} )/{2}$   \\
& $T^ - _{{{i}},{{o}}}$        & $({{T^ \bot _{{{i}},{{o}}} - T^\parallel _{{{i}},{{o}}}}} )/{2}$   \\
& $R^{\perp,\parallel}$          & Fresnel reflection coefficients  \\
& $R^{+}$        & $({{{R^ \bot } + {R^\parallel }}} )/{2}$  \\
& $R^{-}$        & $( {{{R^ \bot } - {R^\parallel }}} ) / {2}$  \\
& $R^{\times}$        & $\sqrt {{R^ \bot }{R^\parallel }}$  \\
 \cline{1-3}
\end{tabular}
}%
\vspace{0mm}%
\end{center}
\end{table}
\section{Conclusion}
\label{sec:conclusion}

Our sparse ellipsometry technique allows us to estimate both shape and polarimetric SVBRDF with high accuracy, yielding results that are on par with full ellipsometry, despite needing only a few minutes (instead of days) of acquisition time, and without the need for complex tabletop capture setups. Instead, we just require a handheld camera setup with fixed polarizing optical elements, with which a series of unstructured photographs are taken. Moreover, our work overcomes the main limitations of two recent related methods that deal with polarization, lifting both the single-view constraint of~\citet{Baek2018}, and the single-material restriction of~\citet{Baek2020}, thus allowing for the first time to capture simultaneously polarimetric SVBRDF and 3D geometry. We expect that our approach will foster the development of novel accurate polarimetric imaging applications with portable devices, not limited to carefully calibrated lab conditions.


\begin{acks}
We thank the reviewers for their valuable feedback that has helped to improve our paper.
We also thank the members of the Graphics and Imaging Lab for proofreading the manuscript.
Min H.~Kim acknowledges the MSIT/IITP of Korea (RS-2022-00155620 and 2017-0-00072) and the Samsung Research Funding Center (SRFC-IT2001-04) for developing partial 3D imaging algorithms, in addition to the support of the NIRCH of Korea (2021A02P02-001), Samsung Electronics, and Microsoft Research Asia.
Diego Gutierrez and Adolfo Muñoz acknowledge funding from the European Research Council (ERC) under the EU's Horizon 2020 research and innovation program (project CHAMELEON, Grant no. 682080), the Spanish Ministry of Science and Innovation (project PID2019-105004GB-I00), and the Govern of Aragon (project BLINDSIGHT).
\end{acks}

\bibliographystyle{ACM-Reference-Format}
\bibliography{bibliography}

\end{document}